\documentclass[twocolumn]{aastex62}

\graphicspath{{./}{figures/}}

\submitjournal{ApJ}

\shorttitle{SN 2018gep}
\shortauthors{Pritchard et al.}
\begin{document}

\title{The Exotic Type Ic Broad-Lined Supernova SN 2018gep:\\
Blurring the Line Between Supernovae and Fast Optical Transients}

\correspondingauthor{T. A. Pritchard}
\email{tapritchard@nyu.edu, tylerapritchard@gmail.com}

\author[0000-0002-0786-7307]{T. A. Pritchard}
\affil{Center for Cosmology and Particle Physics, New York University, 726 Broadway, NY, NY 11201, USA}

\author{Katarzyna Bensch}
\affiliation{Instituto de Astrof\'isica de Andaluc\'ia, IAA-CSIC, Glorieta de la Astronom\'ia, S/N, 18008, Granada, Spain}

\author{Maryam Modjaz}
\affil{Center for Cosmology and Particle Physics, New York University, 726 Broadway, NY, NY 11201, USA}

\author{Marc Williamson}
\affil{Center for Cosmology and Particle Physics, New York University, 726 Broadway, NY, NY 11201, USA}

\author[0000-0002-7978-7648]{Christina C. Th\"one}
\affiliation{Instituto de Astrof\'isica de Andaluc\'ia, IAA-CSIC, Glorieta de la Astronom\'ia, S/N, 18008, Granada, Spain}

\author[0000-0001-8764-7832]{J. Vink\'o}
\affiliation{CSFK Konkoly Observatory, Konkoly Thege M. ut 15-17, Budapest, 1121 Hungary}
\affiliation{Department of Optics \& Quantum Electronics, University of Szeged, D\'om t\'er 9. Szeged, 6720 Hungary}
\affiliation{ELTE E\"otv\"os Lor\'and University, Institute of Physics, P\'azm\'any P\'eter s\'et\'any 1/A, Budapest, 1117 Hungary}
%\email{vinko@konkoly.hu}

\author[0000-0003-1953-8727]{Federica B. Bianco}
\affil{{Department of Physics and Astronomy, University of Delaware, Newark, DE, 19716, USA}}
\affil{{Joseph R. Biden, Jr. School of Public Policy and Administration, University of Delaware, Newark, DE, 19716, USA}}
\affil{{Data Science Institute, University of Delaware, Newark, DE, 19716, USA}}

\author{K. Azalee Boestroem}
\affiliation{Department of Physics, University of California, Davis, CA 95616, USA}

\author{Jamison Burke}
\affiliation{Las Cumbres Observatory, 6740 Cortona Drive, Suite 102, Goleta, CA 93117-5575, USA}
\affiliation{Department of Physics, University of California, Santa Barbara, CA 93106-9530, USA}

\author[0000-0002-7077-308X]{Rub\'en Garc\'ia-Benito}
\affiliation{Instituto de Astrof\'isica de Andaluc\'ia, IAA-CSIC, Glorieta de la Astronom\'ia, S/N, 18008, Granada, Spain}

\author{L. Galbany}
\affiliation{Departamento de F\'isica Te\'orica y del Cosmos, Universidad de Granada, E-18071 Granada, Spain}

\author[0000-0002-1125-9187]{Daichi Hiramatsu}
\affiliation{Las Cumbres Observatory, 6740 Cortona Drive, Suite 102, Goleta, CA 93117-5575, USA}
\affiliation{Department of Physics, University of California, Santa Barbara, CA 93106-9530, USA}

\author[0000-0003-4253-656X]{D. Andrew Howell}
\affiliation{Las Cumbres Observatory, 6740 Cortona Drive, Suite 102, Goleta, CA 93117-5575, USA}
\affiliation{Department of Physics, University of California, Santa Barbara, CA 93106-9530, USA}

\author[0000-0001-9695-8472]{Luca Izzo}
%\affiliation{Instituto de Astrofísica de Andalucía, IAA-CSIC, Glorieta de la Astronomía,s/n,18008, Granada, Spain.}
\affiliation{DARK, Niels Bohr Institute, University of Copenhagen, Lyngbyvej 2, DK-2100 Copenhagen, Denmark}

\author[0000-0003-2902-3583]{D. Alexander Kann}
\affiliation{Instituto de Astrofísica de Andalucía, IAA-CSIC, Glorieta de la Astronomía,s/n,18008, Granada, Spain.}

\author[0000-0001-5807-7893]{Curtis McCully}
\affiliation{Las Cumbres Observatory, 6740 Cortona Drive, Suite 102, Goleta, CA 93117-5575, USA}
\affiliation{Department of Physics, University of California, Santa Barbara, CA 93106-9530, USA}

\author{Craig Pellegrino}
\affiliation{Las Cumbres Observatory, 6740 Cortona Drive, Suite 102, Goleta, CA 93117-5575, USA}
\affiliation{Department of Physics, University of California, Santa Barbara, CA 93106-9530, USA}

\author[0000-0001-7717-5085]{Antonio de Ugarte Postigo}
\affiliation{Instituto de Astrofísica de Andalucía, IAA-CSIC, Glorieta de la Astronomía,s/n,18008, Granada, Spain.}
\affiliation{DARK, Niels Bohr Institute, University of Copenhagen, Lyngbyvej 2, DK-2100 Copenhagen, Denmark}

\author[0000-0001-8818-0795]{Stefano Valenti}
\affiliation{Department of Physics, University of California, Davis, CA 95616, USA}

\author{Xiaofeng Wang}
\affiliation{Physics Department, Tsinghua University, Beijing, 100084, China}

\author{J. C. Wheeler}
\affiliation{Department of Astronomy, University of Texas at Austin, Austin, TX 78712, USA}

\author{Danfeng Xiang }
\affiliation{Physics Department, Tsinghua University, Beijing, 100084, China}

\author{K. S\'arneczky}
\affiliation{CSFK Konkoly Observatory, Konkoly Thege M. ut 15-17, Budapest, 1121 Hungary}
%\email{sarneczky.krisztian@csfk.mta.hu}

\author[0000-0002-8585-4544]{A. B\'odi}
\affiliation{CSFK Konkoly Observatory, Konkoly Thege M. ut 15-17, Budapest, 1121 Hungary}
\affiliation{MTA CSFK Lend\"ulet Near-Field Cosmology Research Group}
\affiliation{ELTE E\"otv\"os Lor\'and University, Institute of Physics, P\'azm\'any P\'eter s\'et\'any 1/A, Budapest, 1117 Hungary}
%\email{bodi.attila@csfk.mta.hu}

\author{B. Cseh}
\affiliation{CSFK Konkoly Observatory, Konkoly Thege M. ut 15-17, Budapest, 1121 Hungary}
%\email{cseh.borbala@csfk.mta.hu}

\author[0000-0003-3759-7616]{D. Tarczay-Neh\'ez}
\affiliation{CSFK Konkoly Observatory, Konkoly Thege M. ut 15-17, Budapest, 1121 Hungary}
\affiliation{MTA CSFK Lend\"ulet Near-Field Cosmology Research Group}
%\email{tarczaynehez.dora@csfk.mta.hu}

\author{L. Kriskovics}
\affiliation{CSFK Konkoly Observatory, Konkoly Thege M. ut 15-17, Budapest, 1121 Hungary}
\affiliation{ELTE E\"otv\"os Lor\'and University, Institute of Physics, P\'azm\'any P\'eter s\'et\'any 1/A, Budapest, 1117 Hungary}
%\email{kriskovics.levente@csfk.mta.hu}

\author{A. Ordasi}
\affiliation{CSFK Konkoly Observatory, Konkoly Thege M. ut 15-17, Budapest, 1121 Hungary}
%\email{ordasi.andras@csfk.mta.hu}

\author{A. P\'al}
\affiliation{CSFK Konkoly Observatory, Konkoly Thege M. ut 15-17, Budapest, 1121 Hungary}
\affiliation{ELTE E\"otv\"os Lor\'and University, Institute of Physics, P\'azm\'any P\'eter s\'et\'any 1/A, Budapest, 1117 Hungary}
%\email{apal@szofi.net}

\author{R. Szak\'ats}
\affiliation{CSFK Konkoly Observatory, Konkoly Thege M. ut 15-17, Budapest, 1121 Hungary}
%\email{szakats.robert@csfk.mta.hu}

\author{K. Vida}
\affiliation{CSFK Konkoly Observatory, Konkoly Thege M. ut 15-17, Budapest, 1121 Hungary}
\affiliation{ELTE E\"otv\"os Lor\'and University, Institute of Physics, P\'azm\'any P\'eter s\'et\'any 1/A, Budapest, 1117 Hungary}
%\email{vida.krisztian@csfk.mta.hu}

%\textbf{}


\begin{abstract}
In the last decade a number of rapidly evolving transients have been discovered that are not easily explained by traditional supernovae models. We present optical and UV data on onee such object, SN 2018gep, that displayed a fast rise with a mostly featureless blue continuum around maximum light, and evolved to develop broad features more typical of a SN Ic-bl while retaining significant amounts of blue flux throughout its observations.  The blue excess is most evident in its near-UV flux that is over 4 magnitudes brighter than other stripped envelope supernovae, but also visible in optical g$-$r colors at early times.  Its fast rise time of $t_{\rm rise,V} \lesssim 6.2 \pm 0.8$ days puts it squarely in the emerging class of Fast Evolving Luminous Transients, or Fast Blue Optical Transients.  With a peak absolute magnitude of M$_r=-19.49 \pm 0.23 $ mag it is on the extreme end of both the rise time and peak magnitude distribution for SNe Ic-bl.  Only one other SN Ic-bl has similar properties, iPTF16asu, for which less of the important early time and UV data have been obtained.  We show that the objects SNe 2018gep and iPTF16asu have similar photometric and spectroscopic properties and that they overall share many similarities with both SNe Ic-bl and Fast Evolving Transients. We obtain IFU observations of the SN 2018gep host galaxy and derive a number of properties for it including  M$_{host}= 7.8^{+2.4}_{-1.2} \times 10^7$\,M$_\odot$ and a metallicity of log(O/H)$+12 = 8.31^{+0.07}_{-0.09}$.  We show that the derived host galaxy properties for both SN 2018gep and iPTF16asu are overall consistent with the SNe Ic-bl and GRB/SNe sample while being on the extreme edge of the observed Fast Evolving Transient sample.  These photometric observations are consistent with a simple SN Ic-bl model that has an additional form of energy injection at early times that drives the observed rapid, blue rise, and we speculate that this additional power source may extrapolate to the broader Fast Evolving Transient sample.   \\
\end{abstract}

%\keywords{choose wisely}


\section{Introduction} \label{sec:intro}
As recent transient surveys have begun to detect an increasing number of transients \citep{Bellm19,Chambers2016,ASAS-SN} due to an increase in both cadence and volume of sky, new types have been discovered as well as outlier objects in otherwise well-understood classes \citep{Kasliwal12}.  Broad-lined Type Ic (Ic-bl) Supernovae (SNe) are a sub-subclass of stripped envelope supernovae (SESNe) that are canonically classified by a lack of H \& He observed in their spectrum \citep[Ic SNe;][]{Filippenko97,GalYam17,Modjaz19-review} and that have an observed Fe velocity of $\gtrsim 1.5  \times 10^4$km s$^{-1}$ \citep{Modjaz16}. While SNe Ic-bl constitute an intrinsically rare class of SNe  \citep[$\sim$ 4\% of the SESN rate\footnote{Note the caveat that this SN Ic-bl rate is based on only one object in the LOSS sample.};][]{Shivvers17}, the overall number of SNe Ic-bl has increased dramatically in the last several years \citep{Bianco14,Modjaz16,Taddia19,Shivvers19}. In general, they have a broader range of light curve rise times, including very rapid rises,
%more rapid rise 
and more luminous peak magnitudes than other SESNe; thus, they have larger inferred $^{56}$Ni masses and explosion energies \citep{Cano13,Taddia15,Prentice16} than other SESNe.  %They are also the only class of SNe that are directly related to long-duration Gamma-Ray Bursts (GRBs) \citep{Woosley06,Modjaz11-rev,Cano17_obs_guide} although not every SN~Ic-bl is observed to be accompanied by a GRB. It has been estimated that at most 85 \% of SNe Ic-bl with radio emission could be off-axis GRBs \citep{Corsi16}.  {\em something about progenitors}, especially if they explode in the same low-metallicity environments as SNe Ic-bl with GRBs \citep{Modjaz20}.
They are also the only class of SNe that are directly connected to long-duration Gamma-Ray Bursts (GRBs) \citep{Woosley06,Modjaz11-rev,Cano17_obs_guide} although not every SN~Ic-bl is observed to be accompanied by a GRB. The question whether SNe Ic-bl without observed GRBs may have produced jets is hotly debated: e.g., while \citet{Corsi16} suggest based on their radio data (mostly upper limits) from a sample of PTF SNe Ic-bl that  less than 85\% of those SNe Ic-bl may have harbored off-axis GRBs (i.e, the GRBs occurred but were not directed toward our line-of-sight), that study assumed densities and GRB energies that only apply to some cosmological GRBs, but are not shared by the most common kind of low-luminosity GRB, such as SN~2006aj/GRB060218. Now a picture is emerging in which the broad lines in SNe Ic-bl may be caused by a jet, even if seen off-axis, as suggested by the hydro plus radiative transfer models in \citet{Barnes18} and as claimed for SN~2020bvc (\citealt{Izzo20_20bvc}, but see \citealt{Ho20_20bvc}), and in which SNe Ic-bl  share the same low-metallicity environments as SN-GRBs \citep{Modjaz20}, and thus the same kind of low-metallicity progenitor.

Rare, known sub-classes of SNe are not the only objects to have been discovered in the ever increasing data volume of transients.  Recent discoveries of optical transients that evolve on the $\sim 1-2$ week timescales with luminosities comparable to that of SNe have been discovered \citep[for recent reviews, see e.g,][]{Inserra19, Modjaz19-review}. Called variously ``Rapidly Evolving Luminous Transients" \citep[RELTs;][]{Drout14}, ``Rapidly Rising Luminous Transients"\citep[][]{Arcavi16},  and ``Fast Evolving Luminous Transients"\citep{Rest18}, ``Rapidly Evolving Transients" \citep[RETs;][]{Pirsiainen2018}, and ``Fast Blue Optical Transients" \citep[FBOTs;][]{Inserra19}. They are an inhomogeneously observed class of objects whose progenitor systems and explosion mechanisms are unknown. The variety of names reflects the variety observed across the samples - some transients \citep[e.g.][]{Arcavi16, Pirsiainen2018} have a variety of colors and are not strictly blue but do evolve rapidly. Some samples consist strictly of more luminous objects \citep{Arcavi16} while others have a broader range of luminosities \citep{Drout14, Pirsiainen2018}.  Potential explanations for these transient events have included magnetar powered explosions, an explosive shock running into dense circum-stellar medium (CSM), off-axis GRB afterglows, black-hole formation in a failed supernovae and the birth of binary neutron star systems. Studies suggest that they are not {\em intrinsically} rare, with a rate of $\sim 5-10$ \% of the Core-Collapse SN Rate \citep{Drout14}, but that the detection efficiency in most transient surveys are low due to these transients being sparsely sampled in a $\sim 3$ day cadence.    

We present here observations of SN~2018gep, which was spectroscopically identified as a SNe Ic-bl by discovery teams (\ref{sec:disc}), but as we show, exhibits some features that are different from those of SNe Ic-bl and similar to those of rapidly evolving transients. In Section (\ref{sec:obs}) we discuss our photometric and spectroscopic observations of this object. In Section (\ref{sec:lc}) we discuss its photometric properties in comparison to others in the class of SN Ic-bl and others in similar regions of the transient rise-time vs peak magnitude parameter space. In Section (\ref{sec:spectra}) we examine our spectra of SN 2018gep and compare them to those of other objects. In Section (\ref{sec:host}) we discuss our spectroscopic long-slit and IFU studies of its host galaxy. In Section (\ref{sec:discussion}) we discuss the implications of SN 2018gep for understanding both SNe Ic-bl and Fast Evolving Transients.  

\section{Observations} \label{sec:obs}
  \subsection{Discovery \& Classification}\label{sec:disc}
\begin{figure}[ht!]
   \centering 
   \includegraphics[width=5.5cm]{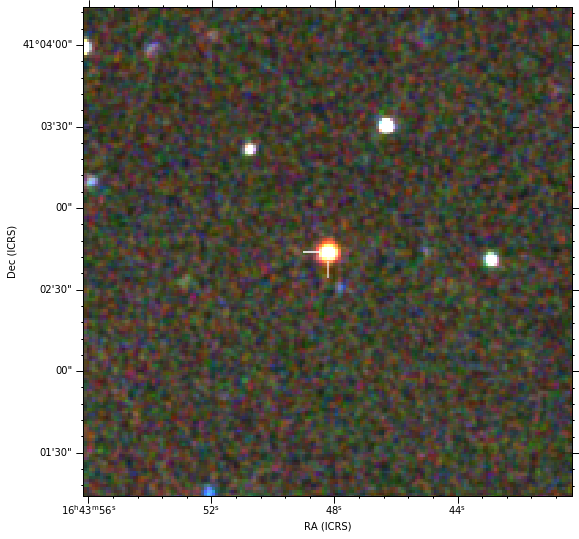} 
   \includegraphics[width=5.5cm]{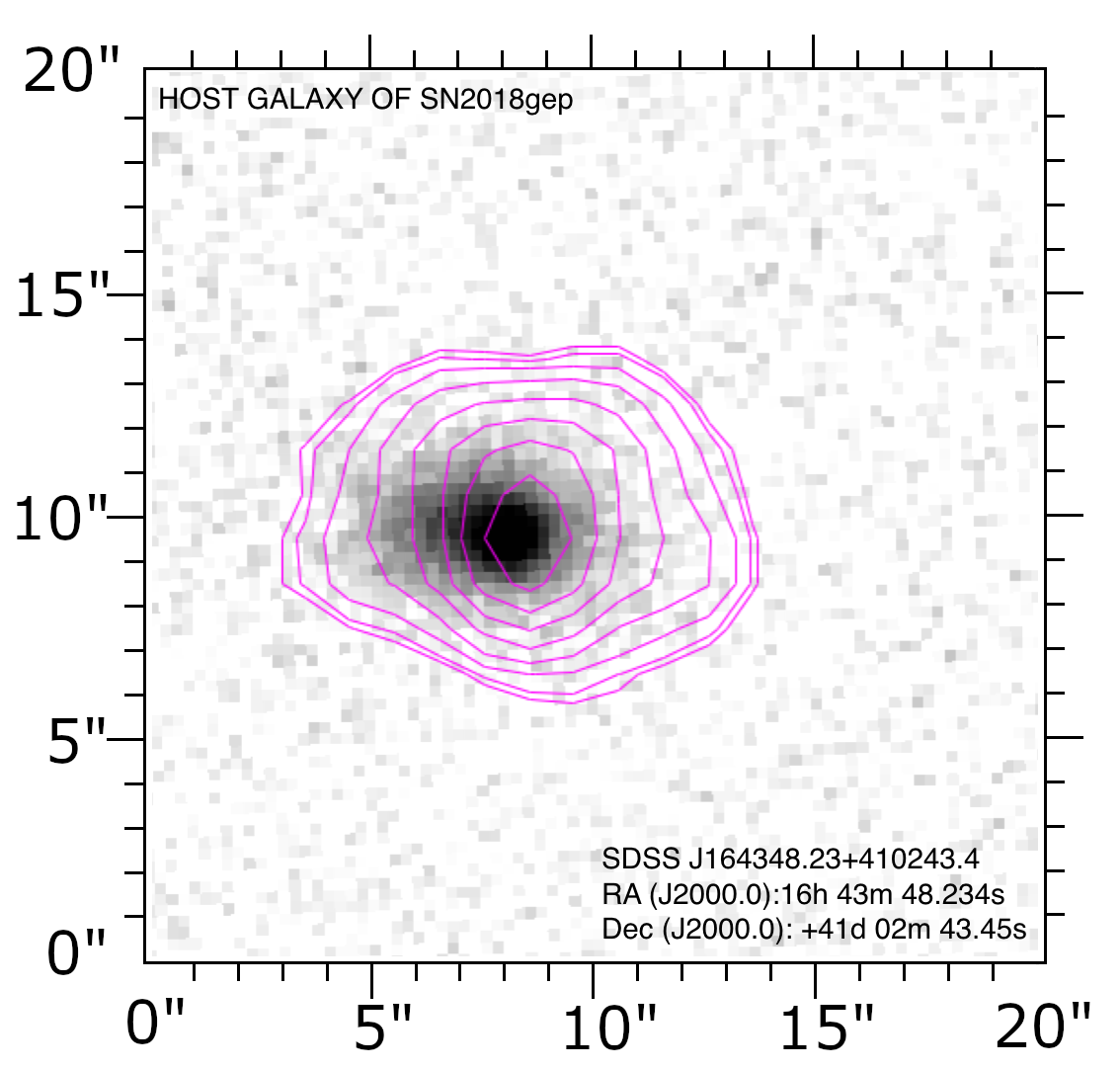}
   \caption{{\em Top: } {\em Swift} u/b/v composite color image of SN 2018gep and its host galaxy around maximum light.  {\em Bottom:} PanSTARRS-1 $g^\prime$ image with contours that are equivalent to a $g^\prime$-band image that we extracted from the IFU data - see Section \ref{sec:host} for more details.}
  \label{fig:PPAKFOV}
  \end{figure}

  SN 2018gep/ZTF18abukavn (Figure \ref{fig:PPAKFOV}, {\em Top}) was first discovered on 03:55:17 09 September 2018 (JD$=2458370.6634$) by \citet{TNS22559} as part of the public ZTF survey  \citep{Bellm19} at (RA, Dec) = (16:43:48.22, +41:02:43.37).  %With a discovery magnitude of 20.48 in the r-ZTF filter and a previous upper limit from half an hour earlier at  03:22:45 it was identified as a young, early transient.  \\ % Something is weird with this double check
  Approximately ten days later on 19 September 2018, \citet{TNS2804}, as part of the Global Supernovae Project (GSP), obtained an optical spectrum (see Section \ref{sec:spec}) and classified the object as a broad-line Type Ic supernovae (Ic-bl) with an ejecta velocity of $\sim 24000$ km/s and a redshift of 0.032 which is consistent with the probable host galaxy identified by \citet{TNS22559}, SDSS J164348.22+410243.3 with a $z=$ 0.033 with a SN-host separation of $\sim 1.5\arcsec$.
  
\subsection{Photometry}
\begin{figure}[ht!]
\includegraphics[width=0.49\textwidth]{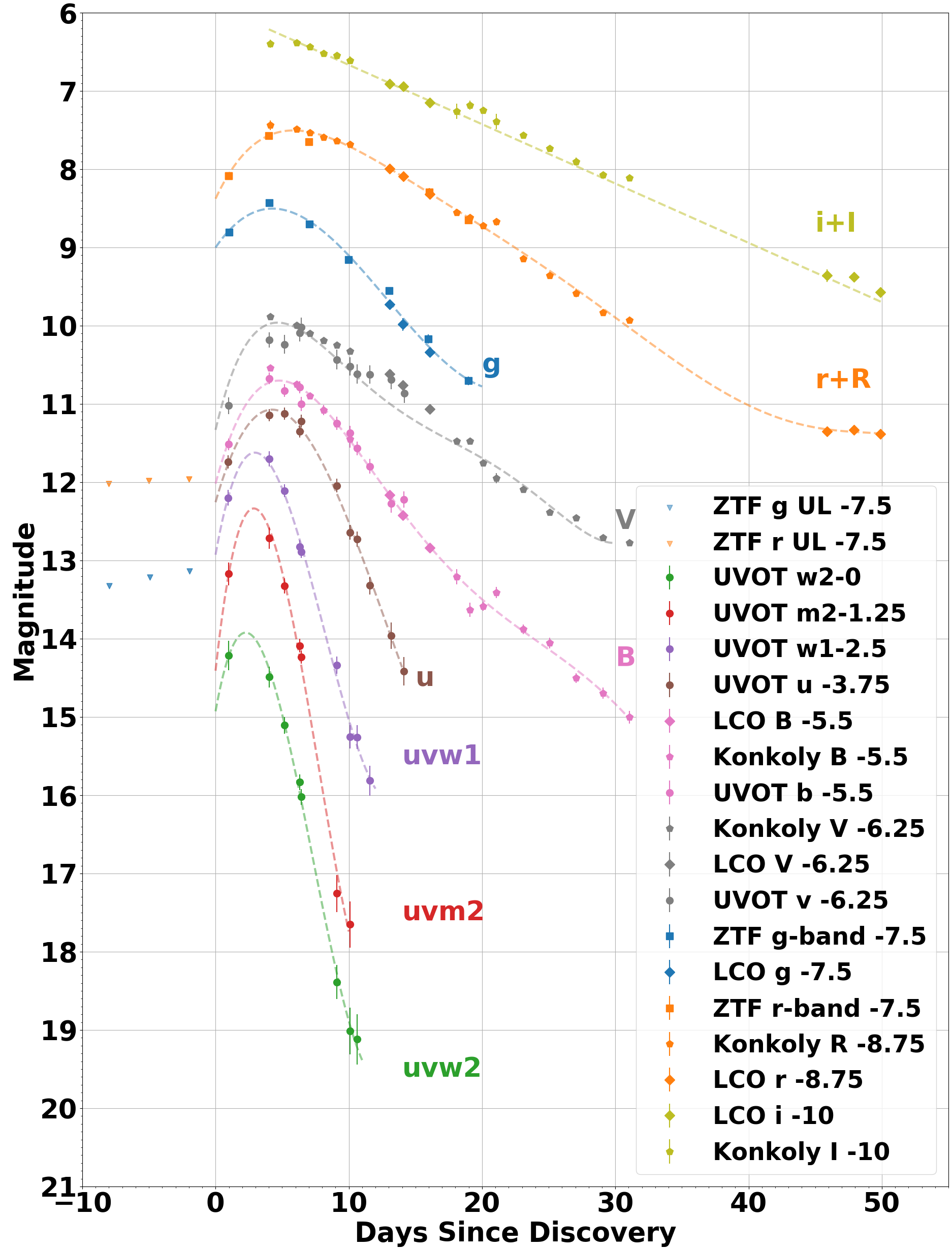}
\caption{Multi-color photometry (and upper limits) of SN~2018gep. The lines are low order polynomial lines fit to the data  purely for visual clarity.    \label{fig:lc}}
\end{figure}

The ZTF public survey observed SN~2018gep between 08 September 2018 and 28 September 2018 in the r-ZTF and g-ZTF filters.  ZTF data were obtained from public alerts made available by the  Las Cumbres Observatory MARS broker that provides access to the publicly available background subtracted ZTF data products.\\  
The Global Supernovae Project (GSP) obtained additional Las Cumbres Observatory (LCO) BVgri-band follow-up data with the Sinistro and Spectral cameras on 1m and 2m telescopes, respectively. Using lcogtsnpipe \citep{Valenti16}, a PyRAF-based photometric reduction pipeline, PSF fitting was performed. Reference images were obtained with the Sinistro and Spectral Imager after the SN faded and image subtraction was performed using PyZOGY \citep{Guevel17}, an implementation in Python of the subtraction algorithm described in \citet{Zackay16}. BV-band data were calibrated to Vega magnitudes using the AAVSO Photometric All-Sky Survey \citep[APASS,][]{Henden09}, while gri-band data were calibrated to AB magnitudes using the Sloan Digital Sky Survey \citep[SDSS,][]{SDSSDR15}.  Science observations were taken between 22 September 2018 and 30 October 2018 with template photometry taken between 22-26 January 2019.   \\
Additional photometric observations were collected with the 0.6/0.9m Schmidt telescope at Piszkesteto Mountain Station of Konkoly Observatory, Hungary, using the 4k$\times$4k FLI CCD equipped with Johnson-Cousins-Bessel $BVRI$ filters. After the usual bias-, dark- and flatfield corrections, PSF photometry was performed on the SN and a set of nearby stars used as tertiary standards. Photometric calibration was done using PS1 photometry on the local tertiary standards, after transforming the catalogued $g_P, r_P, i_P$ magnitudes to $BVRI$ ones via the calibration by  \citet{Tonry12}. Finally, the flux contribution from the host galaxy was taken into account by computing aperture photometry on the host as appeared on the PS1 frames and subtracting its fluxes from the ones obtained from PSF-photometry on the Konkoly frames. \\
Observations with the {\em Neil Gehrels Swift Observatory} \citep{Gehrels04} Ultra-Violet/Optical Telescope \citep[UVOT,][]{Roming05} began on 14:02:56 09 September 2018 ($\sim 0.5$ days after discovery) using three optical (u, b, v) and three UV filters \citep[uvw2, uvm2, uvw1: $\lambda_c=1928$, 2246, 2600 {\AA} respectively;][]{Poole08} after being triggered by \citet{ho2019}. Regular observations  continued through 03 October 2018 with a final observation obtained 29 October 2018.  Data were reduced using the process described in \citet{Pritchard14} with the final observation used for galaxy template subtraction.  While there may be some small contamination from the supernova at this time, any UV emission is far below {\em Swift} sensitivity at this time frame and the optical observations from {\em Swift} are consistent with the other sources presented here (LCO, Konolly).
Data from these sources are presented in Figure \ref{fig:lc} and made available in Table \ref{tab:PhotObs}.  

\begin{table}[ht!]
\caption{\label{tab:PhotObs}Photometry of SN 2018gep.}
\centering                           
\begin{tabular}{lllll} 
\hline\hline                
  \textbf{JD} &\textbf{mag}&\textbf{mag$_\mathrm{err}$}&\textbf{Instrument}&\textbf{Filter}\\
  \hline
2458383.7278  & 17.665    &    0.0194 & LCO 2m0-01 &  B \\
2458383.7326  &  16.864   &    0.0154 & LCO 2m0-01 &  V \\
2458383.7376  &  17.230   &    0.0089 & LCO 2m0-01 &  g \\
2458383.7412  &  16.741   &    0.0262 & LCO 2m0-01 &  r \\
2458383.7445  &  16.909   &    0.0143 & LCO 2m0-01 &  i \\
 \hline
\end{tabular}
\tablecomments{Table 1 is published in its entirety in the machine-readable format.
      A portion is shown here for guidance regarding its form and content.}
\end{table}

\subsection{Spectroscopy}\label{sec:spec}
We obtained optical spectroscopy of the SN as well as its host galaxy and list the journal of our spectroscopic observations in Table \ref{tab:SpecObs}.

%Breaking on web version; fix - error w/ deluxe tables and non-deluxe tables, recreated below
%\begin{deluxetable*}{llllllll}
%\tablenum{1}
%\tablewidth{0pt}
%\tablecaption{\label{tab:SpecObs}Spectroscopic Observations of SN~2018gep and Its Host galaxy.}
%\tablehead{
%\colhead{UT Date} &
%\coldhead{$t_{\rm V,max}\tablenotemark{a}$} &
%\colhead{Tel. + Instr.} &
%\colhead{Wave. Range} &
%\colhead{P.A.} &
%\colhead{Airmass} &
%\colhead{Slit} &
%\colhead{Exp.} \\
%\colhead{ } &
%\colhead{(days)} &
%\colhead{ } &
%\colhead{(\AA)} &
%\colhead{($^\circ$)} &
%\colhead{ } &
%\colhead{($^{\prime\prime}$)} &
%\colhead{(sec)} }
%\startdata
%2018-09-11.31                  &$-$3.7& \text{OGG 2m+FLOYDS}               & 3700$-$10000  &   95.3  & 1.74   & 2.0 %& 1800\\
%2018-09-19.24                  &$+$4.3& \text{OGG 2m+FLOYDS}               & 3700$-$10000  & 112.6   & 1.30   & 2.0 %& 1800\\
%2018-11-07.45\tablenotemark{b} & $+$ 54.5& \text{CAHA+PMAS}\tablenotemark{c} & 3750$-$7500   & N/A     & 2.2 & N/A & %3x1200 \\
%2019-02-05.64\tablenotemark{b} & $+$144.6& \text{Keck+LRIS}                  &  3200$-$9200  & 250     &1.33& 1.0 & %  900
%\enddata
%\tablenotetext{a}{Days with respect to $V$-band maximum} 
%\tablenotetext{b}{No SN light, only host galaxy}
%\tablenotetext{c}{IFU observations, thus long-slit information such as slit size and P.A. is not applicable here.} 
%\end{deluxetable*}

\begin{table*}[]
    \begin{center}
    \caption{\label{tab:SpecObs}Spectroscopic Observations of SN~2018gep and Its Host galaxy.}
    \begin{tabular}{llllllll}
    \hline\hline                
    \textbf{UT Date} & \textbf{t$_{\rm V,max}$ \tablenotemark{a}} & \textbf{Tel. + Instr.} & \textbf{Wave. Range} & \textbf{P.A.} & \textbf{Airmass} & \textbf{Slit} & \textbf{Exp.} \\
    \textbf{ } & \textbf{(days)} & \textbf{ } & \textbf{(\AA)}  & \textbf{($^\circ$)} & & \textbf{($^{\prime\prime}$)} & \textbf{(sec)} \\
  \hline
2018-09-11.31                  &$-$3.7& \text{OGG 2m+FLOYDS}               & 3700$-$10000  &   95.3  & 1.74   & 2.0 & 1800\\
2018-09-19.24                  &$+$4.3& \text{OGG 2m+FLOYDS}               & 3700$-$10000  & 112.6   & 1.30   & 2.0 & 1800\\
2018-11-07.45\tablenotemark{b} & $+$ 54.5& \text{CAHA+PMAS}$^c$ & 3750$-$7500   & N/A     & 2.2 & N/A & 3x1200 \\
2019-02-05.64\tablenotemark{b} & $+$144.6& \text{Keck+LRIS}                  &  3200$-$9200  & 250     &1.33& 1.0 &   900 \\
\hline
    \end{tabular}
    \end{center}
\tablenotetext{a}{Days with respect to $V$-band maximum} 
\tablenotetext{b}{No SN light, only host galaxy}
\tablenotetext{c}{IFU observations, thus long-slit information such as slit size and P.A. is not applicable here.} 
\end{table*}

%Two optical spectra were obtained by the LCO-GSP on 11 and %19 September 2018 using the OGG 2m telescope with the %FLOYDS spectrograph (R$\sim 400-700$ over $\lambda \sim %3700-10000$\AA) 
%and reduced using the automated LCO pipepline. %{\em %\color{red} Floyds CITE? ; Jamie Burke reduced - sound %good?}.  
%These data are discussed and analyzed in Section %\ref{sec:spectra}. \\ 

Additional observations of the location of SN 2018gep and its host galaxy were obtained by two different telescopes at 1.5-2 months after the explosion. One was obtained via Director's Discretionary Time (PI: Bensch) using the Potsdam MultiAperture Spectrophotometer \citep[PMAS; ][]{Roth05}, which is an Integral Field Unit instrument (IFU), mounted on the 3.5m telescope at the Centro Astron\'omico Hispano en Andaluc\'ia (CAHA). The other was with the Low-Resolution Imaging Spectrometer (LRIS) \citep{Oke95,McCarthy98,Rockosi10} at the 10m W. M. Keck Observatory on Maunakea, Hawaii, as part of the LCO-GSP follow-up program
%to study nebular emission of supernovae 
(PI: Valenti), using a long-slit aperture.

 The IFU observations using the PMAS instrument in PPAK mode \citep{Verheijen2004,Kelz2006} were carried out on 7 November 2018. We used the V500 grating with G$_\mathrm{rot}=143.5$, which covers a wavelength range between $\sim3750-7500$~\AA{} at  a resolution of 6.5~\AA{} FWHM, corresponding to $\sim350$~km\,s$^{-1}$. The PPAK IFU consists of 331 science fibers with diameters of 2\farcs7. The science fibers are placed in a hexagonal parcel resulting in a filling factor of 65\%, and cover a field-of-view (FOV) of $72''\times64''$. For sky subtraction 36 sky fibers are placed around the science fibers. An additional 15 fibers illuminated by internal lamps were used to calibrate the instrument. Three science exposures of 1200\,s each were obtained at a signal-to-noise ratio (SNR) of $\sim10$ per \AA{} for the spectral continuum. 
 %It was only possible to perform 1h exposure per night due to the early setting of this galaxy. 
 We used a dithering pattern consisting of three pointings to cover the entire FOV including the spaces in-between the fibers. In Figure (\ref{fig:PPAKFOV}, {\em Bottom}) we show the FOV of the PPAK IFU and the region around the host which is plotted in the subsequent figures that display host-galaxy properties.

To reduce the PMAS-PPAK data we used a python-based pipeline that executes the following steps: identification of the position of the spectra on the detector along the dispersion axis; extraction of each individual spectrum; distortion correction of the extracted spectra; wavelength calibration; fiber-to-fiber transmission correction; flux-calibration; sky-subtraction; cube reconstruction; and finally differential atmospheric correction \citep[for more details see: ][]{Garcia-Benito10,Husemann13,Garcia-Benito15}. 
These IFU data and their analysis are discussed as part of our host-galaxy study in Section \ref{sec:host}.  

The late-time long-slit Keck spectrum was reduced in the standard way using the LPIPE pipeline \citep{Perley19} - no SN emission was detected at the location of SN~2018gep, neither the 1D nor the 2D spectra and the spectrum is thus included as part of our host-galaxy study in Section \ref{sec:host} and included in the spectroscopic observations table. % {\em \color{red} Azalee Boestrom - Keck Host Details - sound good?  What should I add?}\\
%\footnote{http://www.astro.caltech.edu/ dperley/programs/lpipe.html} 
%\subsection{Late Time emission from 18gep}
%
%{\em \color{blue} Should we say something about the uncertainties in our extraction and %potential Co decay from jozsef's plot but lack of nebular emission?  This whole thing is still %concerning to me. MARYAM says: no \\
%
%Also, throw a table in when all the data is finalized}

\section{Light Curve Analysis} \label{sec:lc}
\subsection{Rise time \& Absolute Magnitude Comparison\label{sec:rt}}
The combined UV-optical lightcurves for SN 2018gep are shown in Figure \ref{fig:lc}. From the {\em Swift} v-band data we calculate that the epoch of maximum light in the v-band is  t$_{peak, V}=58375.7 \pm 0.8$ MJD using the Monte-Carlo method outlined in \citet{Bianco14}. The relatively deep upper limits from the ZTF survey provide for a strong constraint on the rise time - using the last ZTF upper limit as an upper limit on the explosion date we calculate a rise time of t$_{rise, v}\lesssim 6.2 \pm 0.8 $ days. Using this same method we calculate the observed peak magnitude in the r-band, $m_{r,peak}=16 \pm 0.05$ mag. %mmodjaz: Need to finalize this! Also we should mention the value (and how it was obtained) that we ended up using in my review paper 
%You: " Also we should mention the value (and how it was obtained) that we ended up using in my review paper" - chat about this, not sure exactly what you used, this value hasn't really changed I just hadn't put the number in b/c there was a discussion between two number's depending on if we use fed's code or not (which I felt was not a great fit for this object)
%mmodjaz: We used mmodjaz: 18gep/Pritchard,5.9,2,-19.87,0.23 mmodjaz: where the format is sn-name,rise,rise_e,,rabs,rabs_e
From the observed redshift of $z=0.031875\pm0.000075$ (See Section \ref{sec:host}) we calculate the absolute magnitude for SN 2018gep to be M$_{\rm V}$=$-19.47\pm 0.23$ mag using the astropy %what about error bars? In my review I used -19.87 +/- 0.23 which you had sent me at some point   Based on our discussion: include how the peak mag was measured and the uncertainty calculated . Based on our discussion: include how the peak mag was measured and the uncertainty calculated .
\citep{astropy1,astropy2} cosmology package and a flat $\Lambda$CDM model with H$_0$=74.22 km~s$^{-1}$Mpc$^{-1}$ \citep{Reiss19} and $\Omega_m$=0.286.  This cosmology model is used throughout the rest of this work for consistency.

\begin{figure}[ht!]
\includegraphics[width=0.49\textwidth]{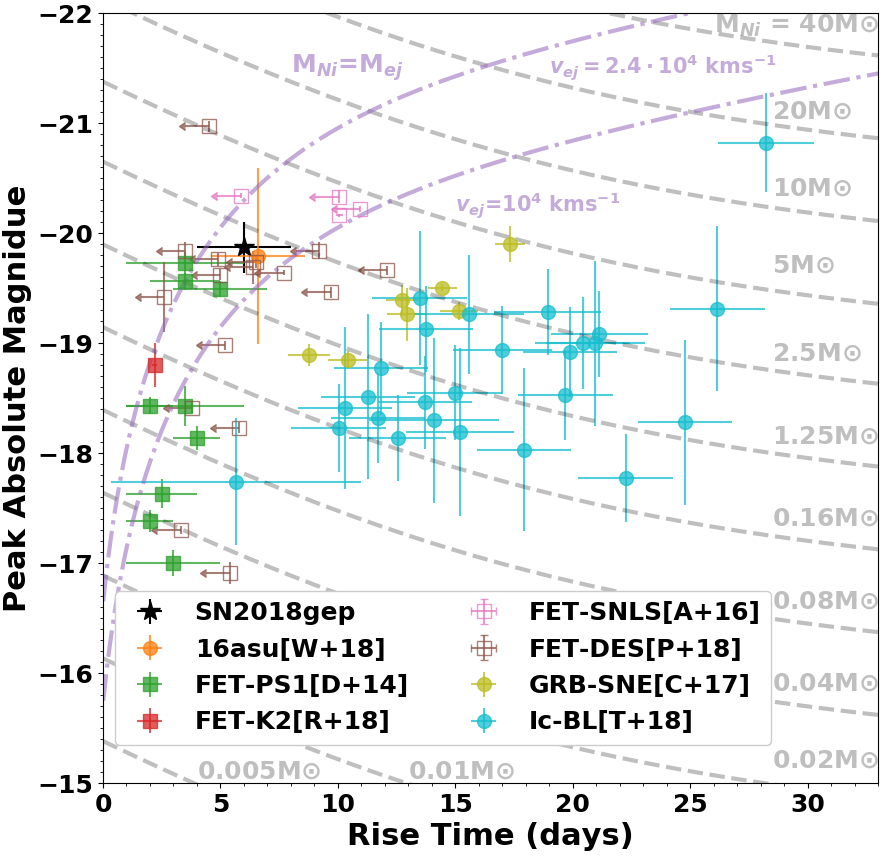}
 \caption{Rise Time vs. Peak Magnitude for a variety of transient sources.  The PTF/iPTF SNe Ic-bl without observed GRBs \citep[T+19;][]{Taddia19} cluster around the GRB/SNe \citep[C+17;][]{Cano17} with a small gap in rise time between most of the sample (PTF10vgv has some overlap, see \ref{fig:lccomp} and the Fast Evolving Transients \citep[FET - D+14,A+16,P+18,R+18; ][]{Drout14,Arcavi16,Pirsiainen2018,Rest18}.  The SN Ic-bl that occupies a similar position as SN~2018gep in this phase space is iPTF16asu \citep[W+18, T+19;][]{Whitesides17,Taddia19}, which is discussed in more detail in Sections \ref{sec:lc}. \label{fig:mvr}}
\end{figure}

Assuming a SN is powered by the typical $^{56}$Ni-decay model, for a particular absolute magnitude and SN rise time, we may calculate an ejecta mass and nickel fraction as outlined in \citet{Arcavi16} \citep[Eqns1\&2 and following from ][]{Arnett82,Stritzinger05,Wheeler15}. It is important to note that this approach makes a number of simplifying assumptions including: spherical symmetry, a constant opacity, a central nickel concentration and that the photospheric velocity is characteristic of the ejecta velocity. This relation is therefore more indicative than strict, and in Figure \ref{fig:mvr} we sketch out lines for a series of ejecta masses and two additional lines corresponding to objects in which the ejecta mass must be entirely composed of nickel to power their light curve. If an objects lies above these lines, an additional source of energy injection or a different source of power is required. Since this limit also depends on ejecta velocity, we draw two lines: the lower line corresponding to a typical SN with $\sim 10,000$ km/s expansion and the top line corresponding to a SN with $\sim 24,000$ km/s expansion velocity as measured from the spectra of SN 2018gep. SN 2018gep, in this parameter space, is like the other luminous fast-rising transients shown, namely right on the border of what can be easily described with simple nickel-powered relations, and it is consistent with being an outlier from the other SNe Ic-bl which are comfortably below this relation. This implies that SN~2018gep most likely had to have an additional powering source besides the decay of $^{56}Ni$ \citep[see also][ for a detailed model involving CSM interaction and pre-explosion mass loss]{ho2019}. \\

\subsection{Light Curve and Color Comparison with other SNe Ic-bl\label{sec:IcLC}}
\begin{figure}[ht!]
\includegraphics[width=0.47\textwidth]{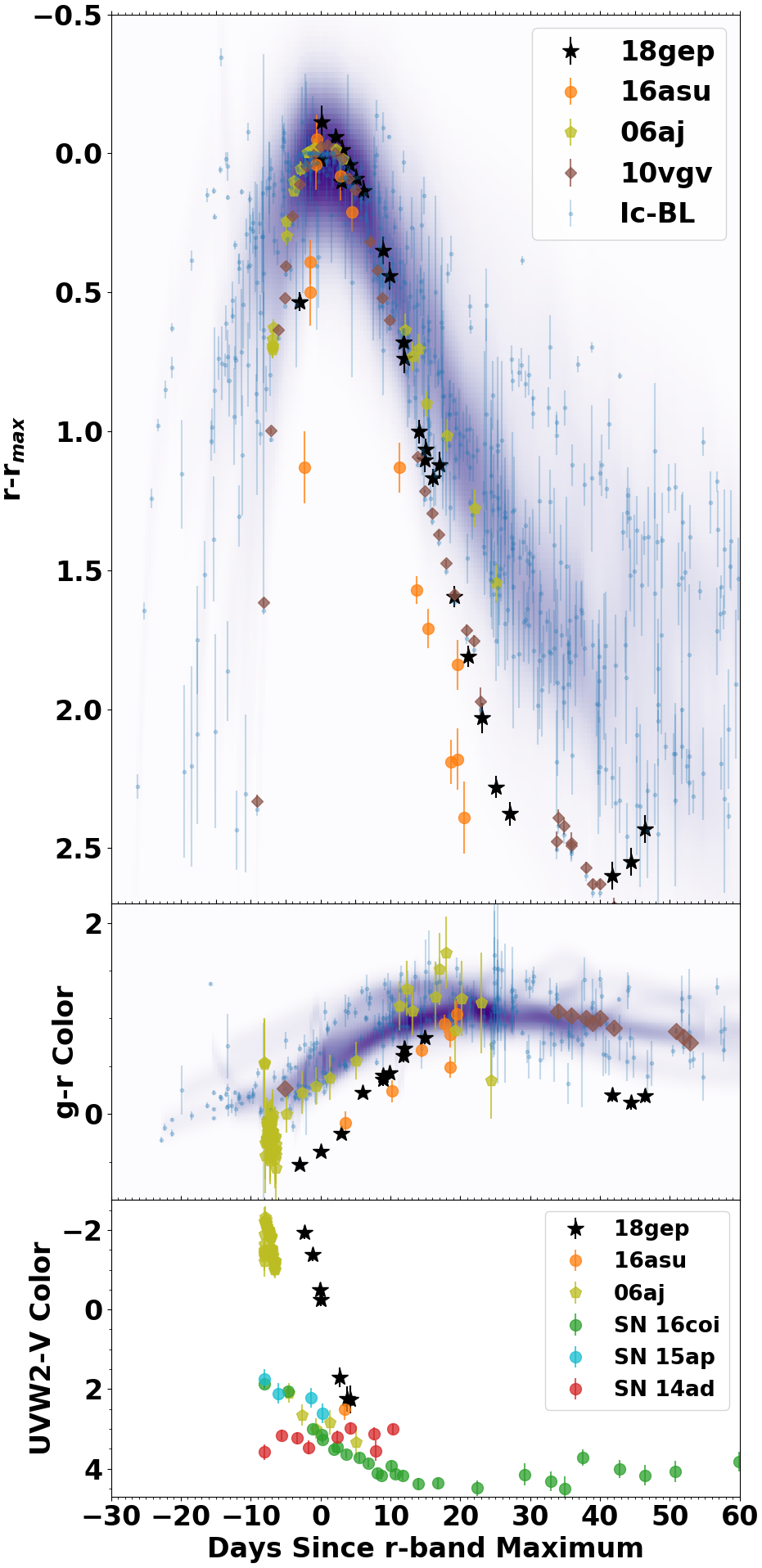}
\caption{Comparison of SN 2018gep with other SNe Ic-bl in the optical \citep[{\em Top, Middle};][]{Taddia19} and UV.  When compared to the \citet{Taddia19} SN Ic-bl sample, the decline rate of SN2018gep is similar to the fastest in that sample (including iPTF16asu), and is significantly more blue at early times than the rest of the sample.  This is even more apparent when we compare the UV emission observed with {\em Swift}, and the only other Ic-bl SNe with similar emission is the early time Shock Cooling (or GRB) emission from GRB 060218/SN 2006aj \citep{Campana06}.  At late times the observed colors of SN~2018gep return to the blue side of the standard SN Ic-bl distribution.     \label{fig:lccomp}}
\end{figure}

In Figure \ref{fig:lccomp} we compare the lightcurve of SN 2018gep with a sample of SNe Ic-bl from \citet{Taddia19} in the optical and a few select SESNe with {\em Swift} UV observations. %give names and references - actually one of them 15ap doesn't seem to be a SN Ic-bl but a SN Ib - check out my long comment on the figure caption for details.  clarify that 06aj may have switched types, cite the appropriate atels
The optical evolution of SN 2018gep is broadly similar to the most rapidly evolving SNe Ic-bl.  The most similar objects are iPTF16asu (an outlier as noted in \citealt{Taddia19}), PTF10vgv and SN 2006aj.     
 While the optical evolution of SN 2018gep and iPTF16asu is similar to that of the SN Ic-bl population as a whole, the early color evolution is {\em not}, particularly in the UV and bluer filters.  As we show in the middle and bottom panels of Figure \ref{fig:lccomp}, there is significantly more blue emission from these SNe at early times than from the rest of the Ic-bl sample, by more than a magnitude in the optical and almost 4 magnitudes in the UV.  By $\sim 10$ days after r-band maximum the color curve of SN 2018gep becomes similar to that of the sample as a whole, albeit remaining somewhat on the blue side.  {\em Only SN 2018gep and iPTF16asu show this significant early blue excess.}  iPTF10vgv and SN~2006aj have similar colors to the SNe Ic-bl sample as a whole.  The only other similarly blue emission is that from SN 2006aj at early times, but it has a significantly faster evolution.  The mechanism that drives this is still a topic of some debate \citep[see][for a discussion]{Irwin16}, and the SNe component of GRB/SNe quickly returns to the `typical' behavior of the SNe Ic-bl sample as a whole - including some phases where SN~2018gep is still UV bright.

%The similarity to iPTF16asu (as noted in section \ref{sec:rt}) holds remarkably well in the light curve comparison as well, where when scaled to the peak magnitude, 18gep rises somewhat more slowly but the decay across both g- and r-bands is quite similar.  These two fast evolving Ic-BL SNe decay significantly slower than other observed Ic-BL SNe, both in the \citet{Taddia19} sample where iPTF16asu was a noted outlier in behavior and in this small comparison with 16coi, a typical Ic-BL with that was well observed in both the UV and optical bands and GRB/SNe 2006aj which is a typical SNe associated with a peculiar GRB afterglow{\color{magenta} cites}.  We also compare this with the super-luminous supernovae 2017ens {\color{magenta} T.W. Chen+} which was a fast evolving and suggested to have some interaction but still evolves on significantly longer timescales.  While SN 2018gep and iPTF16asu are on the bright tail of the distribution of the Fast Blue Optical Transients identified in \citet{Drout14}, the sparse data of the brightest object appears to behave approximately similarlily to SN 2018gep and iPTF16asu, given that the filter comparison is approximate with FBOT ps11qr being observed in the {\color {magenta} u-band at redshift z$\sim$0.5}.  \\

%SN 2018gep was initially quite UV bright, with a strong UV component observed by {\em Swift}.  When the observed data is compared with the small sample of Ic-BL SNe with strong UV defections (Fig \ref{fig:lccomp}, middle) we note that SN 2018gep is almost three magnitudes blueer at early times than any of the other {\em Swift} observed Ic-BL SNe.  In fact, only the GRB afterglow portion of GRB/SNe 2006ag is more blue than SN 2018gep,   Shortly after peak magnitude, however, the UV component becomes negligable and the colors appear to be similar to that of other Type Ic-BL SNe.  

%If we look at optical colors instead (Fig \ref{fig:lccomp}, bottom), we see a similar trend with SN 2018gep being more blue than other Ic-BL at early times, but quite similar to them by $\sim 5$ days after r-band maximum.  With the move to optical, however, we can include data from iPTF16asu and the PanSTARRS FBOTs (with data filters and redshifts matched to most closely approximate the given colors). While the data is approximate, particularly for the FBOTs, it is generally indicative that these objects are more similar to SN 2018gep than other observed Ic-BL SNe.  The sample from \citet{Taddia19} has not been made publicly available, but again iPTF16asu was a noted outlier and given the similar behavior of SN 2018gep it is reasonable to draw a similar conclusion for this object.  

%\begin{figure*}[ht!]
%\includegraphics[width=1\textwidth]{18gep_spectplot.png}
%\caption{SN 2018gep spectra - re-work figure with Marc's %comparison\label{fig:spectcomp}}
%\end{figure*}
\subsection{Comparison with PS1 Fast-Evolving Transients}
\begin{figure}[ht!]
\includegraphics[width=0.37\textwidth]{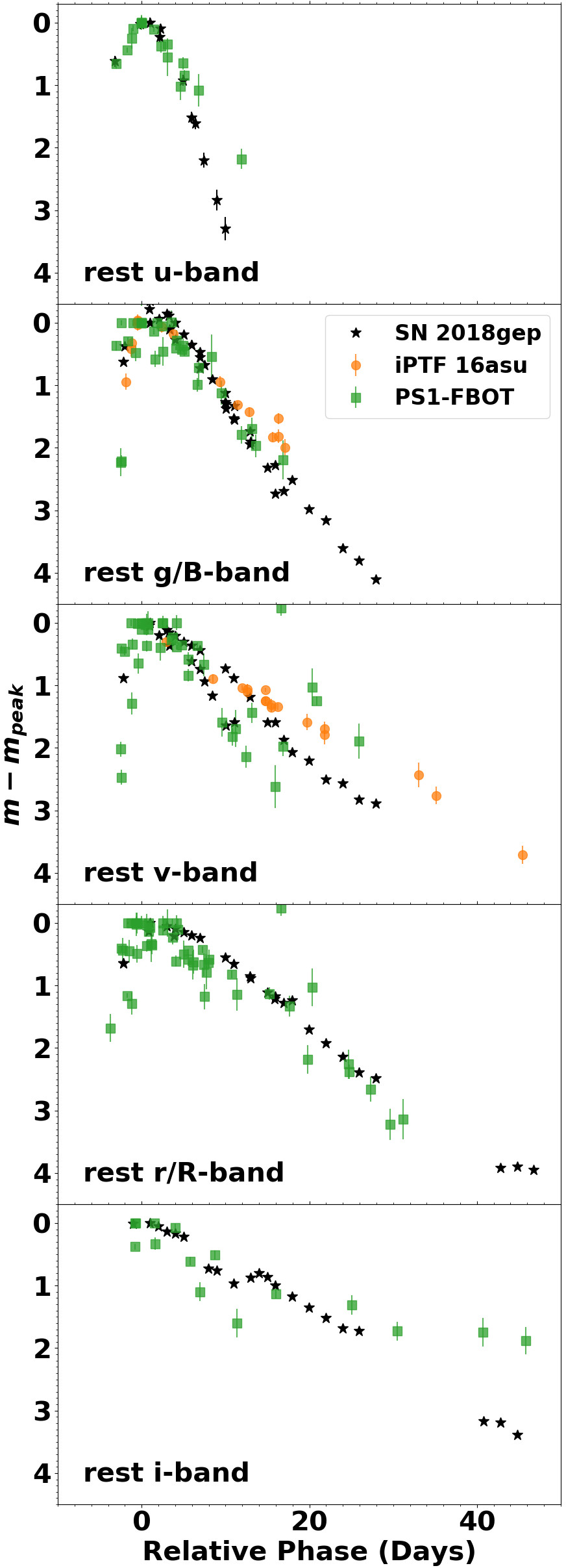}
\caption{Comparison of SN 2018gep with PS1 Fast Evolving Transients from \citet{Drout14}.  Filters have been matched by using the closest rest-frame central wavelength with time dilation but no k-corrections have been applied, implying a qualitative comparison only.  Given the differing band passes and spectral coverage the overall light curve shape between the Fast Evolving Transients and SN 2018gep is quite similar, although some significant scatter may be see in rest V-band around 15-20 days and some deviation at late times in some objects rest u-band and rest i-band. \label{fig:ps1comp}}
\end{figure}

In Sections \ref{sec:rt} \& \ref{sec:IcLC} we show that while SN 2018gep shares similarities with other Ic-bl SNe, it is a notable outlier in terms of color, absolute magnitude, and rise time.  In section \ref{sec:rt} we show that other objects that may behave similarly  to SN 2018gep are the recently discovered class of Fast Evolving Transients first noted by \citet{Drout14} and later in \citet{Arcavi16, Rest18, Pirsiainen2018}.  With many of these objects having poorly constrained rise times due to their rapid evolution, we focus on a comparison with the PS1 sample from \citet{Drout14} which has a significant number of objects with a detected rise as well as multi-color observations.  These are, however, found at a significantly large range of redshifts - to compare we match the observed SN 2018gep band with the closest rest-frame band of a PS1 object, as shown in Figure \ref{fig:ps1comp}.  This is a rather coarse measurement, as the relative filter band-passes are different and a more detailed analysis would perform k-corrections to address this.  However, we choose to avoid k-corrections as they are SED-dependent and we have limited information about the SED's of all PS1 objects, while we know that they undergo significant color evolution.  

As seen in Figure \ref{fig:ps1comp}, both SN 2018gep and iPTF16asu have similar relative light curve shapes as the PS1 fast evolving transient population as whole. Furthermore, the observed g-r colors are similar to the sample as reported in \citet{Drout14}.  There is some suggestion that there may be some longer lived emission in some PS1 Fast transients (as seen in the late-time rest frame i-band comparison and u-band comparison),  although these late time deviations each come from a single PS1 object and it is not clear how homogeneous of a sample these objects are.  

\section{Spectroscopic Analysis}\label{sec:spectra}
\begin{figure}[ht!]
\includegraphics[width=\columnwidth]{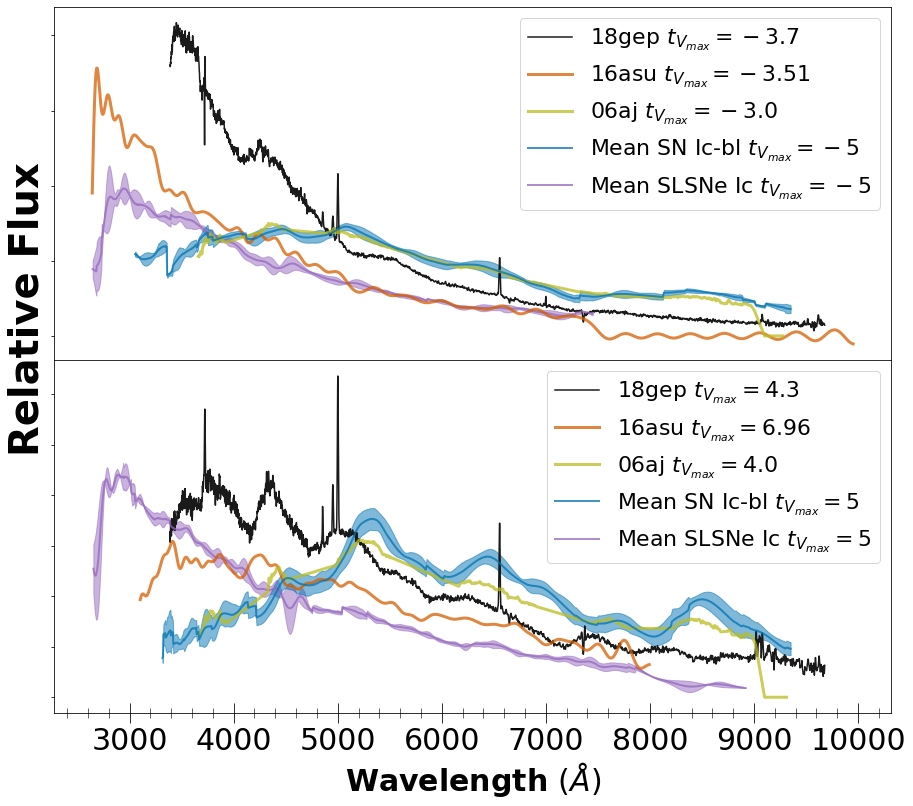}
\caption{Comparison of SN 2018gep (black) spectra to mean (plus standard deviation) spectra of SNe Ic-bl (blue) and SLSNe Ic (purple) classes, along with direct comparisons to SN~2006aj (yellow) and iPTF16asu (orange). The excess blue flux in the SN~2018gep spectra compared to the mean SNe Ic-bl and even SLSNe Ic spectra, and that of iPTF16asu, is clearly evident.\label{fig:meanspec}}
\end{figure}

The spectra of SNe are crucial diagnostics which reveal the elemental composition and dynamics of the ejecta. Since there are relatively few FBOTS with spectra, here we present a detailed analysis of our two spectra of SN~2018gep and their comparison to SN population spectra as well as to individual SN spectra. Our two medium-resolution optical spectra of SN 2018gep at phases $t_{V_{max}}=-3.7$ and $t_{V_{max}}=4.3$ days relative to V-band maximum are shown in Figure \ref{fig:meanspec}. The early spectrum, taken just before maximum light, is characterized by a strong, featureless blue continuum. The later spectrum at phase $t_{V_{max}}=4.3$ days displays broad features typical of an SN Ic-bl spectrum.\footnote{The SN 2018gep spectra have very narrow H$\alpha$ and H$\beta$ emission peaks, which are clearly due to the host galaxy spectrum.} For the post-maximum spectrum, we calculate the absorption and line-width velocities for the FeII $5169$ \AA absorption feature using the techniques from \citet{Modjaz16} and find an absorption velocity $v_{abs}=23800\pm2200$~km/s and a width velocity $v_{lw}=10100^{+300}_{-500}$~km/s. This high absorption velocity is consistent with SN Ic-bl events with associated gamma ray bursts \citep[see Fig. 7;][]{Modjaz16}. 

%Although SN 2018gep has broad spectral features similar to those found in a typical SN Ic-bl spectrum, its light curve and color evolution are better matched by FBOTs. 
In order to evaluate the spectroscopic similarities between SN~2018gep and other SNe, we used the SNID code \citep{blondin2007determining} to match SN~2018gep to other stripped-envelope SNe, whose SNID templates have been produced by \cite{liu2016analyzing}, \cite{Modjaz16}, \cite{liu2017analyzing}, and \cite{williamson2019optimal}. Table \ref{tab:snid} shows the top 5 SNID matches for the $t_{V_{max}}=4.3$ days spectrum of SN 2018gep. SNID cannot match the earliest spectrum due to the lack of supernova features. The majority of the SNID matches are SN Ic-bl spectra, but SNID calculates matches on the continuum-removed spectra. Therefore, the SNID matches only reflect spectral behavior in the absorption lines. In order to investigate the behavior of the continuum in SN~2018gep, we overplot in Figure \ref{fig:meanspec} the mean spectra of SNe Ic-bl \citep[from][]{Modjaz16} and those of Superluminous SNe (SLSNe) from \cite{liu2017analyzing}. SLSNe are included here since they also show broad lines in their spectra \citep{liu2016analyzing, quimby2018spectra}, have blue colors ( see \cite{Inserra19} for a recent review), and are also suggested to be driven by CSM or magnetars, as we also do for SN~2018gep (see Section \ref{sec:discussion}). In addition, we include the individual objects SN~2006aj \citep{modjaz06,modjaz2014optical} and iPTF16asu \citep{Whitesides17} since they have some similarities to SN~2018gep. At early times, SN 2018gep is distinguished from both SNe Ic-bl and SLSNe Ic spectra by its strong blue continuum. In addition, we can see clearly from the spectra that SN~2018gep is even bluer than iPTF16asu -- especially pre-maximum -- something that could not be discerned from the photometry given the lack of pre-maximum g-band and Swift data for iPTF16asu (note that the y-axis uses relative flux, so differences in color manifest as differences in the overall shape and slope of the spectra). At later times ($t_{V_{max}}=4.3$ days), SN~2018gep resembles the mean SN Ic-bl spectrum and SN~2006aj spectrum for $\lambda>5000$~\AA, but there is clear excess flux in the blue part of the SN~2018gep spectrum, which is consistent with our analysis of the SN~2018gep light curve in Figure \ref{fig:lccomp}. At wavelengths $\lambda<5000$~\AA, the lines in SN~2018gep closely resemble those in iPTF16asu, but its continuum is bluer than that of iPTF16asu. This blue flux excess could be due to interaction with CSM \citep{ho2019}. The color of SN~2018gep is more similar to the color of SLSNe Ic after maximum than it was pre-maximum. 

In summary, our detailed spectral analysis shows that SN~2018gep has lines very similar to those in SNe Ic-bl (in terms of absorption and width velocities), but a much bluer continuum than SNe Ic-bl and iPTF16asu, both before maximum light and after maximum light. In addition, before maximum light, SN~2018gep's spectrum appears to be even bluer than the mean spectrum of SLSNe. 

%commented out - numbering error deulxe & non-deluxe tables
%\begin{deluxetable}{CCC}[ht!]
%\tablecaption{\label{tab:snid}SNID matches to SN 2018gep}
%\tablecolumns{3}
%\tablenum{1}
%\tablewidth{\columnwidth}
%\tablehead{
%\colhead{\hspace{0.75cm}SN}\hspace{0.75cm} &
%\colhead{\hspace{0.75cm}Phase (days)}\hspace{0.75cm} &
%\colhead{\hspace{0.75cm}Classification} }
%\startdata
%\text{2006aj} & -0.2 & \text{Ic-bl} \\
%\text{2003bg} & -19.1 & \text{IIb-pec} \\
%\text{2007uy} & -6.3 & \text{Ib-pec} \\
%\text{2016coi} & -10.6 & \text{Ic-bl} \\
%\text{2006aj} & 5.0 & \text{Ic-bl} 
%\enddata
%\end{deluxetable}

\begin{table}[ht!]
    \begin{center}
    \caption{\label{tab:snid}SNID matches to SN 2018gep at $t_{V_{max}}=4.3$ days}
    \begin{tabular}{lll}
    \hline
    \hline
    \textbf{SN} & \textbf{Phase (days)} & \textbf{Classification} \\
    \hline
    \text{2006aj} & -0.2 & \text{Ic-bl} \\
    \text{2003bg} & -19.1 & \text{IIb-pec} \\
    \text{2007uy} & -6.3 & \text{Ib-pec} \\
    \text{2016coi} & -10.6 & \text{Ic-bl} \\
    \text{2006aj} & 5.0 & \text{Ic-bl} \\
    \hline
    \end{tabular}
    \end{center}
    \tablecomments{The top 5 SNID matches to the $t_{V_{max}}=4.3$ days spectrum for SN 2018gep. Phase is measured relative to the date of V-band maximum. Both SN~2003bg and SN~2007uy exhibited broad lines at early times, in particular during their listed phases, which then disappeared over time \citep{mazzali2009sn, modjaz2014optical}. Thus, these two SNe are called peculiar for their type.}
\end{table}

\section{Host Galaxy Analysis}\label{sec:host}
Here we analyze in detail the host galaxy of SN~2018gep and compare it to those of other SN samples (including well-understood ones) and the general population of star-forming galaxies in order to understand its explosion conditions and progenitor.

The study of the transient's host galaxy environments in order to constrain the progenitor of the particular transient has a rich history \citep[e.g.,][for a review see  \citealt{Anderson2015}]{Modjaz08,Thoene19,Modjaz20}, and is an emerging field for the new kind of transients being discovered by innovative surveys, such as FBOTs. Historically this has been done with longslit spectroscopy, however recent advances in the instrumentation of Integral-Field Units (IFUs) and large samples of nearby SNe from ongoing surveys have allowed these studies to be done with IFUs to enable for increased resolution around the SN site and better resolution of the host galaxy and its assosciated dynamics \citep[see ][for a more general discussion across SNe sub-types]{Kuncarayakti13a,Kuncarayakti13b,Kuncarayakti18,Galbany14a,Galbany16,Galbany18}. 
 
\subsection {IFU data}
This study represents the first IFU host-galaxy study of a fast evolving transient. The PPAK IFU spaxels in our final cube have an angular size of $1''\times1''$, however, the seeing during observations was only 1\farcs8, hence the nominal spatial resolution is lower. For our spatially-resolved analysis of the host galaxy we use custom-written IDL codes to extract emission-line maps and properties from the data cubes. 

\subsubsection{Emission-line analysis}\label{sec:eline}
\begin{figure}[ht!]
   \centering 
      \includegraphics[width=\columnwidth]{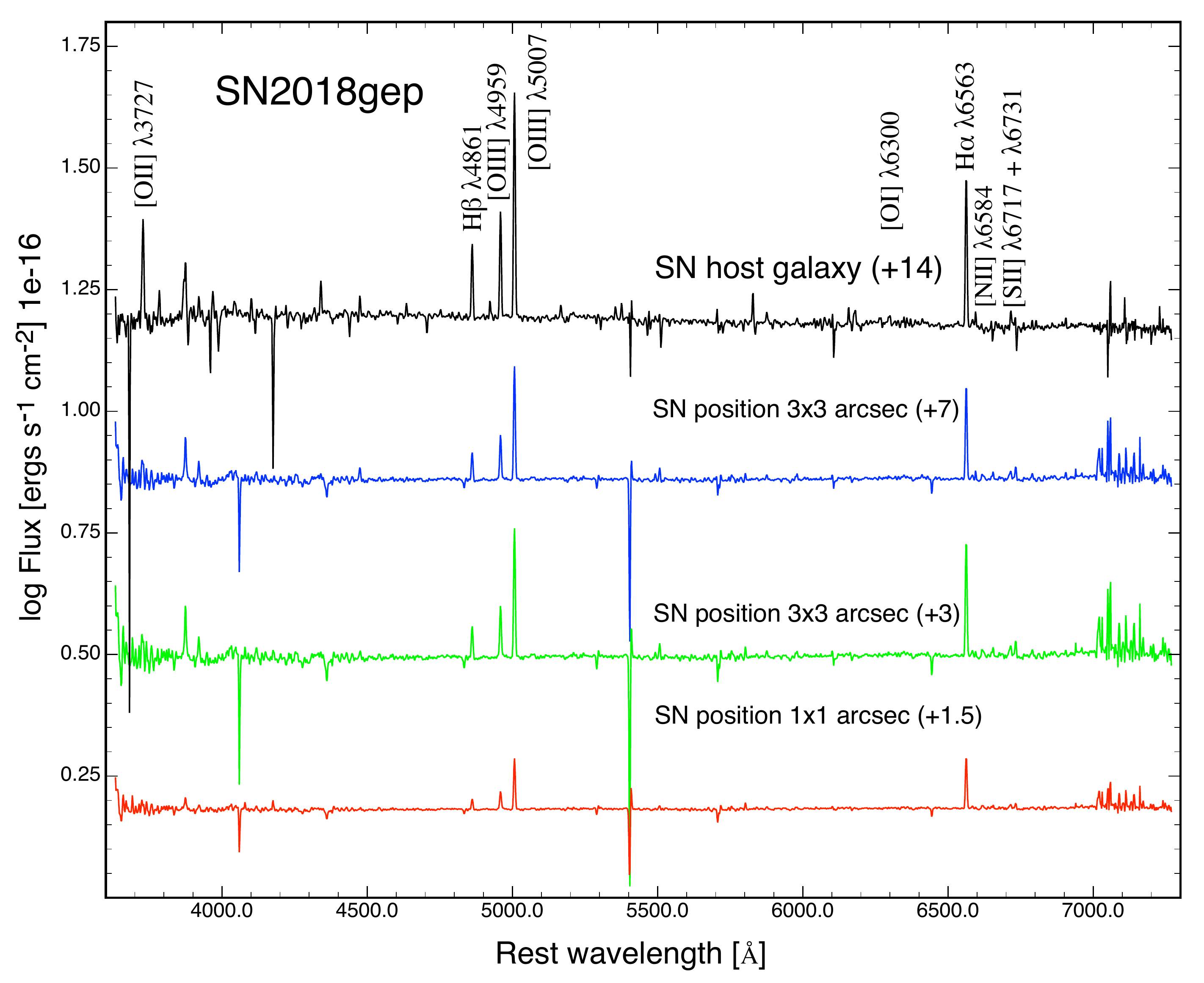}
   \caption{Integrated spectra extracted from the PMAS data cube: the entire galaxy (black) and regions around the SN position using an area of  $1''\times1''$ (red), $3''\times3''$ (five spaxels, green) and $3''\times3''$ (nine spaxels, blue) and are offset for readability.  The strong, narrow absorption lines are residuals from sky line subtraction.}
  \label{fig:extractedIFUspec}
  \end{figure}

In order to obtain emission-line fluxes in each spaxel we sum the fluxes in the spectral direction around the red-shifted position of each emission line and subtract the galaxy continuum. 2D maps of the main emission lines are shown in the Appendix, Figure~\ref{fig:linemaps}. To study the properties of the region around the SN at different spatial resolutions, we extract 1D spectra from 1, 5, and 7 spaxels centered on the SN position using \texttt{QFitsView}\footnote{ http://www.mpe.mpg.de/$\sim$ott/QFitsView}. The spectra are shown in Figure~(\ref{fig:extractedIFUspec}).

%While we will present analysis based on the the full dataset (ie maps showing all spaxels), 

Using the integrated spectrum of the host galaxy we determine a precise redshift from the strong emission lines of H$\beta$ $\lambda$4861\AA, $[$OIII$]$ $\lambda$4959\AA, $[$OIII$]$ $\lambda$5007\AA, $[$OI$]$ $\lambda$6300\AA, H$\alpha$ $\lambda$6563\AA. The mean value obtained from all emission lines yields $z=0.031875\pm0.000075$.
%(we did not applied the heliocentric correction). 
%We corrected the spectra for redshift using the DOPCOR package in IRAF (Doug 1993 \& Doug 1986). 
%We do not detect any stellar absorption We examined each spectrum for stellar absorption using H$\beta$ $\lambda$4861\AA\ line profile and find no sign of it.
%; we compared those extractions with the integrated data from the entire host galaxy and found no detectable supernovae emission. 

To obtain the interstellar extinction in the host galaxy, we use the Balmer decrement of H$\alpha$/H$\beta$ according to \citet{Dominguez2013} adopting the \citet{Calzetti2000} attenuation curve with $R_V=4.05$  which assumes a starburst attenuation law. We assume the standard recombination model for star-forming galaxies and Case B for HI recombination lines \citep{OsterbrockFerland2006}. The intrinsic Balmer decrement at an electron temperature $T=10^4$ K and density $n_e=10^2$ cm$^{-3}$, is expected to be  $j_{H\alpha}/j_{H\beta}=2.86.$   %Finally, we calculated the reddening 
%using the $\textnormal{H}\alpha/\textnormal{H}\beta$ ratio emission line flux ratio \citep{Osterbrock06}.  
%following the equation: 
%$$E(B-V)=1.97\log(\frac{\textnormal{H}\alpha/\textnormal{H}\beta}{2.86}).$$ 
The values obtained for the reddening in the galaxy and the regions around the SN are listed in Table \ref{tab:emissionlines}. The distribution of the extinction across the galaxy is shown in Figure~\ref{fig:SFRZmap}. Curiously, the spectra of the SN line-of-sight region indicate some extinction while the extinction based on the integrated spectrum of the host is consistent with zero. 
As we increase the aperture of extraction from the SN position (see Table \ref{tab:emissionlines}) we see the calculated extinction drop until for the total host-galaxy integrated IFU spectrum the overall extinction is low and consistent with E(B--V)$=0$ mag.  This is also consistent with the zero to low value obtained from the Keck LRIS host-galaxy spectrum.  
One explanation for this apparent discrepancy may be that the overall galaxy emission has little extinction, and while extinction is present throughout the galaxy it is not homogeneously distributed. Therefore the integrated spectrum is dominated of regions with little extinction but more emission, explaining the overall low E(B--V). This is reflected in the IFU map in Figure~\ref{fig:SFRZmap} which shows that the distribution of the extinction is not uniform.   
We observe that the high extinction in the SN line-of-sight region could either imply a considerable amount of dust at the SN site or it is dust behind the SN. 
The latter is favoured by the fact that the SN is observed to be UV bright, $\sim$4 magnitudes bluer in the UV than other Ic-BL at early times. 
If we would de-redden the SN data using extinction values based on the the $1\arcsec\times1\arcsec$  section around the SN in the IFU map ($\textnormal{E(B-V)}=0.5-0.6$ mag, $\textnormal{A}_V=1.8$ mag) the intrinsic peak luminosity in the UV would be unreasonably large.  
%{\bf CT: did we have an explanation for this?? like the light from low-E(B-V) regions dominate or something?? The Keck spectrum also gives zero extinction, so this seems to be consistent...}
%I.e. light weighted

The emission-line fluxes of the spectra are measured using \texttt{SPLOT} in IRAF. Statistical errors were calculated following \citet{Perez-Montero2003}. We found an offset between the SDSS photometry and the magnitude derived from the integrated spectra of $\textnormal{m}-\textnormal{m}_0=0.26$ mag and therefore calibrate the emission-line fluxes using the SDSS $g^\prime$ and $r^\prime$ filters. Fluxes were corrected for Galactic extinction ($\textnormal{A}_\textnormal{v}=0.0286$ mag), and extinction in the host galaxy as estimated in each corresponding spectrum. We list the final extinction-corrected, SDSS-calibrated emission-line values as extracted for different parts of the galaxy in Table~\ref{tab:emissionlines} and Appendix section~\ref{IFU}.

 %We show the distribution of the emission-line fluxes in the host galaxy in the figure \ref{fig:spectcomp}.

\subsubsection{Derived Host properties}
\begin{figure*}[ht!]
   \centering 
      \includegraphics[width=5.7cm]{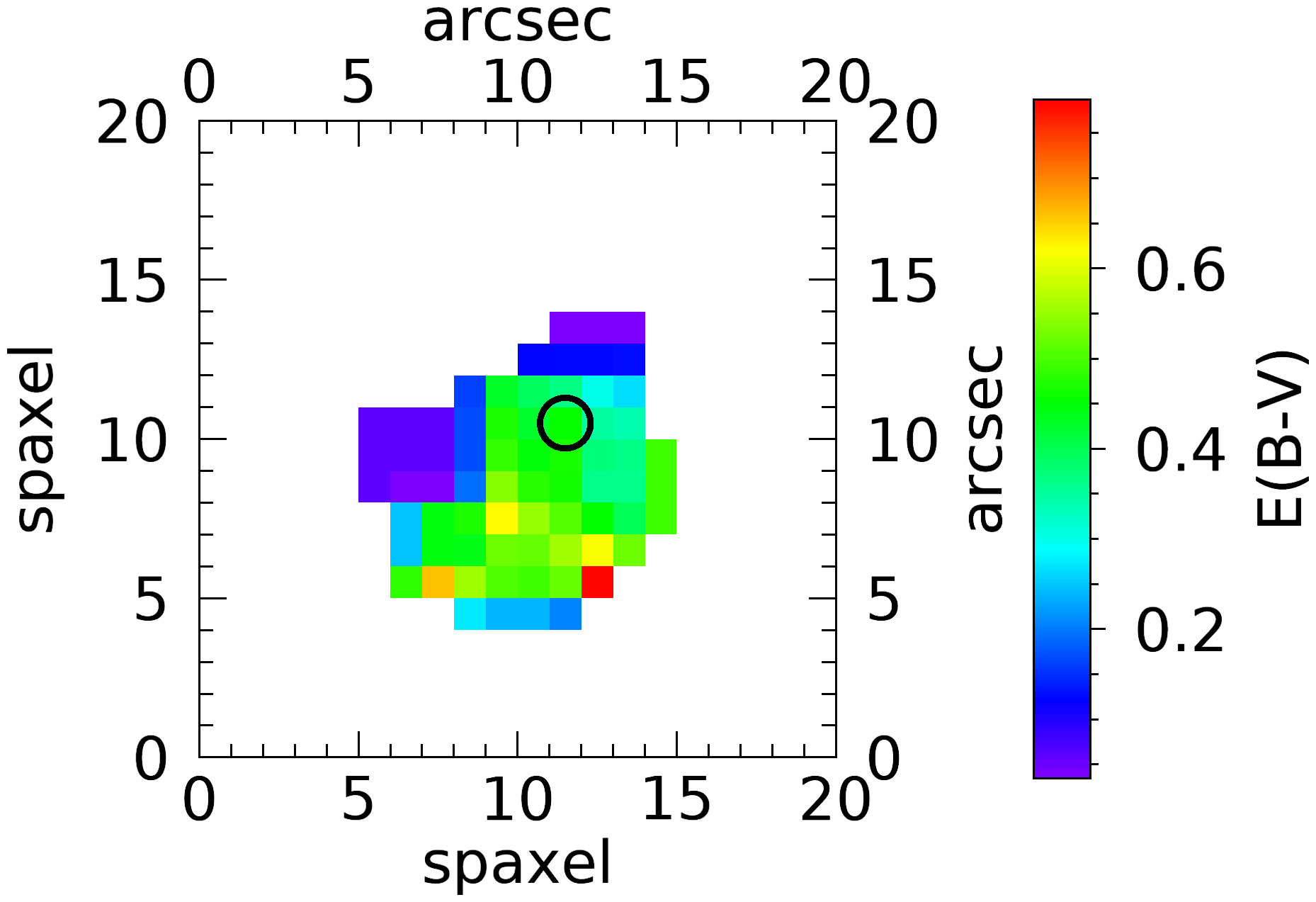}
      \includegraphics[width=5.7cm]{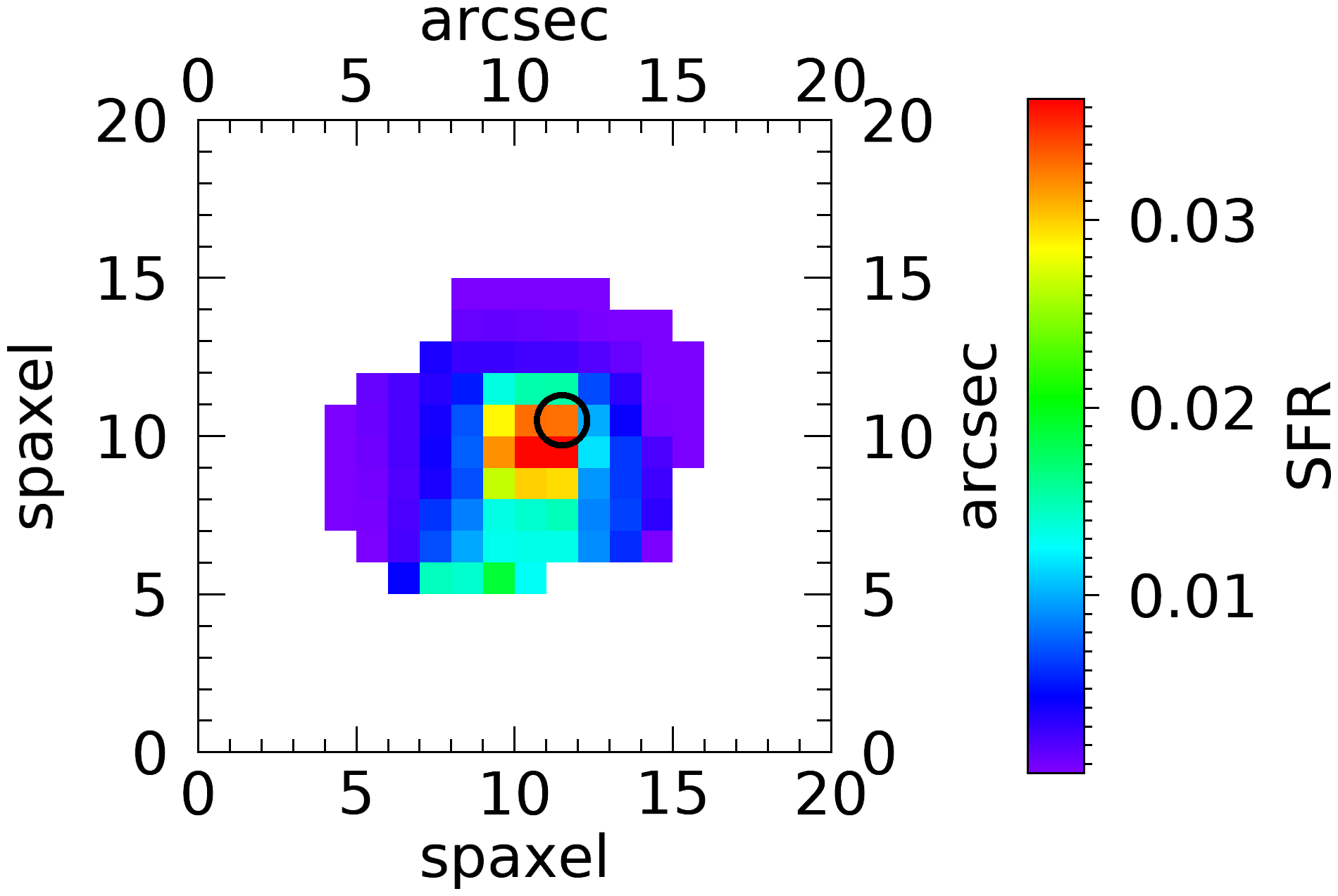}
      \includegraphics[width=5.7cm]{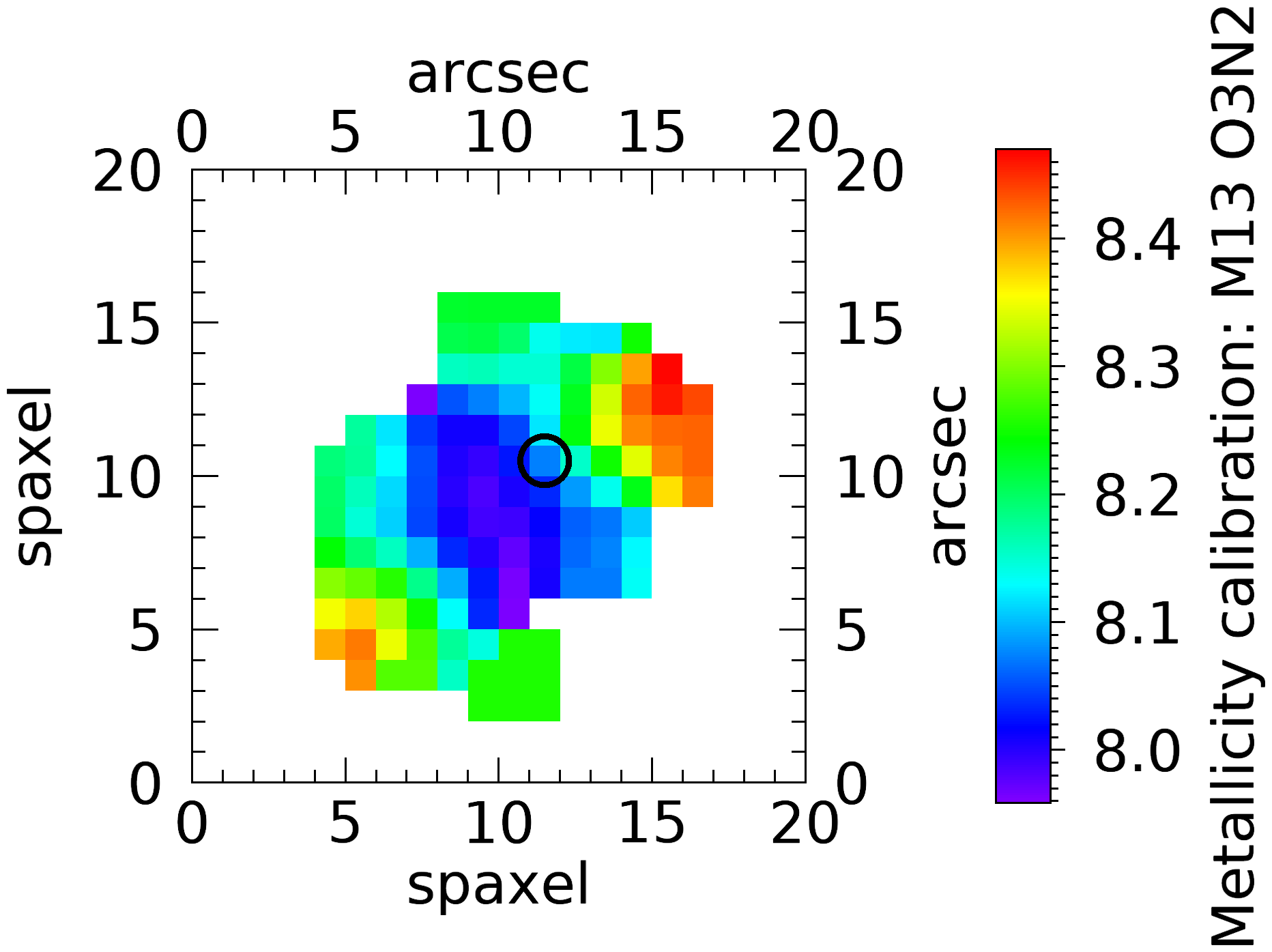}
   \caption{Maps of extinction (left), SFR (middle) and metallicity (right) using the O3N2 parameter in the calibration of \cite{M13O3N2}. The black circle indicates the position of SN 2018gep.}
  \label{fig:SFRZmap}
  \end{figure*}

 The luminosity of the H$\alpha$ nebular line serves as the tracer of the star-formation rate (SFR). To calculate the SFR we follow the relations in \cite{Kennicutt1994} assuming $T=10^4$K and Case B recombination. The values of L(H$\alpha$) and the SFR for both the host galaxy and the SN region are listed in Table~\ref{tab:emissionlines}. The SFR distribution in the galaxy is shown in Figure~\ref{fig:SFRZmap}. 

To determine metallicities (Z) we use the Python code \texttt{pyMCZ} \citep{Bianco2016}, which calculates oxygen abundances using strong-emission-line standard metallicity diagnostics based on a Monte Carlo method to derive the statistical oxygen abundance confidence region. Various emission-line ratios are used in up to 15 theoretical/empirical/combined metallicity calibrations implemented in the code. We present the combination of the emission lines used in each calibration and the results in Table~\ref{tab:metallicity} and refer the reader to the references listed in Table~\ref{tab:metallicity} for a more detailed discussion on the individual diagnostics. Due to its low S/N ratio we decided to exclude $[$OII$]$ $\lambda$3727\AA\ from the metallicity measurements. Our results show no significant difference between the metallicity of the SN region and the integrated host galaxy value.

 \begin{table*}[ht!]
\caption{Emission-line fluxes corrected for Galactic and host-galaxy extinction, and calibrated with SDSS photometry. All fluxes are in 10$^{-16}$  erg s$^{-1}$ cm$^{-2}$.}
\label{FLUX} 
\centering                           
\begin{scriptsize}
\begin{tabular}{lllllllllllllll} 
\hline\hline                
  \textbf{Emission line} &\textbf{$\lambda$ [\AA]}  & \textbf{Host galaxy}  &\textbf{$\lambda$ [\AA]}  & \textbf{SN region} &\textbf{$\lambda$ [\AA]}  & \textbf{SN region}   &\textbf{$\lambda$ [\AA]    }  &  \textbf{SN region} \\
 \textbf{ } &\textbf{ }  & \textbf{}  &\textbf{ }  & \textbf{$1''\times1''$} &\textbf{ }  & \textbf{$3''\times3''$}   &\textbf{ }  &  \textbf{$3''\times3''$} \\
 \textbf{ } &\textbf{ }  & \textbf{ }  &\textbf{ }  & \textbf{ } &\textbf{ }  & (five spaxels)   &\textbf{ }  &  (nine spaxels)\\
  \hline
% \textbf{E(B-V):} & &\textbf{Negative}  & &\textbf{0.493$\pm$0.040} && \textbf{0.403$\pm$0.030}   & & \textbf{0.246$\pm$0.030} \\
% {E(B-V) - \texttt{pyMCZ}:} & &{0.000}  & &\textbf{$0.580_{-0.045}^{+0.052}$} && \textbf{$0.476_{-0.037}^{+0.036}$} & & \textbf{$0.291_{-0.036}^{+0.035}$} \\
% \hline
$[$OII$]$ $\lambda$3727\AA\      &3727.929  &    57.705$\pm$1.249  & 3725.427  &    7.208$\pm$2.241     &  3725.927  &  23.161$\pm$4.500    &   3726.198  &19.215$\pm$3.995   \\ 
H$\beta$ $\lambda$4861\AA\       &4861.477  &    36.445$\pm$0.547  & 4861.556  &    3.649$\pm$0.640     &  4861.404  &  14.952$\pm$1.952    &   4861.405  &16.402$\pm$2.172   \\ 
$[$OIII$]$ $\lambda$4959\AA\     &4959.094  &    54.767$\pm$0.612  & 4959.239  &    6.105$\pm$1.030     &  4959.112  &  25.487$\pm$3.214    &   4959.088  &26.438$\pm$3.385   \\ 
$[$OIII$]$ $\lambda$5007\AA\     &5007.047  &   155.614$\pm$0.894  & 5007.136  &   16.767$\pm$2.773     &  5007.057  &  73.183$\pm$9.054    &   5007.027  &76.130$\pm$9.565   \\ 
$[$OI$]$ $\lambda$6300\AA\       &6297.685  &     8.126$\pm$0.704  & -         &     -                  &  -         &   -                  &   -         &    -       \\        
H$\alpha$ $\lambda$6563\AA\      &6562.696  &    84.348$\pm$1.171  & 6562.640  &   10.326$\pm$1.274     &  6562.660  &  42.315$\pm$3.954    &   6562.650  &46.437$\pm$4.413   \\ 
$[$NII$]$ $\lambda$6584\AA\      &6584.315  &     3.892$\pm$0.584  & 6584.007  &    0.556$\pm$0.096     &  6583.500  &   1.760$\pm$0.309    &   6583.506  & 2.070$\pm$0.445   \\ 
$[$SII$]$ $\lambda$6717\AA\      &6716.741  &     7.075$\pm$0.552  & 6714.937  &    1.790$\pm$0.252     &  6714.704  &   4.303$\pm$0.554    &   6714.525  & 5.982$\pm$0.766   \\ 
$[$SII$]$ $\lambda$6731\AA\      &6732.335  &     5.710$\pm$0.573  & 6732.570  &    1.107$\pm$0.161     &  6732.847  &   4.210$\pm$0.505    &   6732.899  & 5.581$\pm$0.676   \\ 
%$[$SIII$]$ $\lambda$9069\AA\     &  - &  nan     $\pm$    0.0  & - &  nan    $\pm$ 0.0      & - &  nan   $\pm$  0.0       & - &  nan    $\pm$  0.0     \\  
%$[$SIII$]$ $\lambda$9532\AA\     &  - &  nan     $\pm$    0.0  & - &  nan    $\pm$ 0.0      & - &  nan   $\pm$  0.0       & - &  nan    $\pm$  0.0     \\   
 \hline
 \textbf{E(B-V) [mag]:} & &\textbf{0.000}  & &\textbf{0.493$\pm$0.040} && \textbf{0.403$\pm$0.030}   & & \textbf{0.246$\pm$0.030} \\
% {E(B-V) - \texttt{pyMCZ}:} & &{0.000}  & &\textbf{$0.580_{-0.045}^{+0.052}$} && \textbf{$0.476_{-0.037}^{+0.036}$} & & \textbf{$0.291_{-0.036}^{+0.035}$} \\
% \hline
 \textbf{SFR [M$_{\odot}$\ yr$^{-1}$]:} && \textbf{0.139}   && \textbf{0.017}   && \textbf{0.070}   && \textbf{0.076}  \\ 
 %\textbf{L [10$^{6}$ L$_{\odot}$]:} && \textbf{5.224}   && \textbf{0.640}   && \textbf{2.621}   && \textbf{2.876}  \\
 \hline
\end{tabular}
%$[$OIII$]$ $\lambda$4363\AA\ not visible \vfill
\end{scriptsize}
\label{tab:emissionlines}
\end{table*}

Fig.~\ref{fig:SFRZmap} shows distributions of metallicities across the galaxy using the  calibration of \cite{M13O3N2}. Metallicities for other calibrators are shown in the Appendix section~\ref{IFU} for comparison. 

\begin{table*}[ht!]
\caption{Oxygen abundances}
\label{tab:metallicity} 
\centering                           
\begin{scriptsize}
\begin{tabular}{c c c c c c c c c} 
\hline\hline                
\textbf{Calibrator}	& \textbf{Support lines}	& \textbf{Host galaxy}  & \textbf{SN region} & \textbf{SN region }   &  \textbf{SN region} \\
				& 						&  \textbf{ }  & \textbf{$1''\times1''$} & \textbf{$3''\times3''$}   &  \textbf{$3''\times3''$} \\
				& 						&  \textbf{ }  & \textbf{} & (five spaxels)   &  (nine spaxels)\\
\hline%newresults
D02$^{1}$              &     N2                                            &   8.15   +  0.15    -  0.15&    8.20   +  0.15   -  0.16  &    8.11   +  0.16   -  0.16 &  8.13   +  0.16   -  0.17 \\
PP04 N2H$\alpha^{2}$ &     N2                                              &   8.15   +  0.02    -  0.03&    8.17   +  0.03   -  0.04  &    8.13   +  0.03   -  0.03 &  8.14   +  0.04   -  0.04 \\
PP04 O3N2$^{2}$        &     N2, O3/H$\beta$                               &   8.10   +  0.02    -  0.02&    8.11   +  0.04   -  0.04  &    8.07   +  0.03   -  0.04 &  8.08   +  0.04   -  0.04 \\
M08 N2H$\alpha^{3}$   &     N2                                             &   8.24   +  0.05    -  0.07&    8.30   +  0.07   -  0.08  &    8.20   +  0.07   -  0.08 &  8.22   +  0.08   -  0.09 \\
M13 O3N2$^{4}$         &     [N II]$\lambda$6584/H$\beta$, O3/H$\beta$                &                 &    8.18   +  0.01   -  0.01  &    8.18   +  0.00   -  0.00 &  8.18   +  0.01   -  0.01 \\
M13 N2$^{4}$           &     [N II]$\lambda$6584/H$\beta$                  &   8.13   +  0.05    -  0.05&    8.16   +  0.06   -  0.06  &    8.11   +  0.06   -  0.06 &  8.12   +  0.06   -  0.06 \\
KK04 N2$\alpha^{5}$  &     N2, q, (N2O2)                                   &   8.26   +  0.05    -  0.07&    8.31   +  0.08   -  0.09  &    8.21   +  0.07   -  0.08 &  8.23   +  0.09   -  0.10 \\
KD02comb$^{6}$         &     COMBINED$^*$                                  &   8.26   +  0.05    -  0.07&    8.31   +  0.07   -  0.09  &    8.21   +  0.07   -  0.08 &  8.23   +  0.09   -  0.10 \\ 
\hline
\end{tabular}
\\$^*$This method chooses the optimal among given: M91, KD02 N2O2, KD02 N2Ha, KD04 R23, [N2, N2O2] diagnostics \citep{Kewley08}\\
References: 
$^1$ \cite{D02}
$^2$ \cite{PP04}
$^3$ \cite{Maiolino08}
$^4$ \cite{M13O3N2}
$^5$ \cite{KK04}
$^6$ \cite{Kewley08}
\end{scriptsize}
\end{table*}

\subsection{Host longslit spectroscopy}
We also obtained one long-slit spectrum of the host using LRIS/Keck. The LRIS spectrum is a light-weighted average of a $1\arcsec\times4\arcsec$ size region centered on the ``nucleus" of the galaxy (i.e. the one with the strongest trace/continuum). 
%{\bf CT: What is the size of the slit and how was this one placed across the galaxy? Hence, what do we actually probe? KB: Maryam? I have no such info.} 
Fluxes were measured using SPLOT in IRAF and errors calculated in the same way as for the integrated regions from the PMAS data. The fluxes are presented in  Table~\ref{tab:KECKfluxes} in the Appendix. We corrected all fluxes for Galactic extinction ($\textnormal{A}_V=0.0286$ mag). We determine the intrinsic extinction using the Balmer decrement as described above and found no extinction based on this spectrum. This result is consistent with the value of the extinction based on IFU integrated galaxy spectrum, but is not consistent with the extinction deduced from IFU data at the SN position, which indicates a large Balmer decrement in that region. We only see high extinction at the SN region as we explain in Section \ref{sec:eline}. where we speculated that it may be due to dust that is accumulated in a small area behind the SN. Hence, without clear emission lines, the extracted LRIS spectrum with area of $1\arcsec\times4\arcsec$ centered on the galaxy ``nucleus", may miss some light from the SN region.

In the Keck spectrum we detect the same lines as in the integrated IFU spectrum%{\color{red}MM - LRIS has high S/N 3727 which we dont trust in IFU?}
, and additionally we measure the [SIII] lines at $\lambda9069$ and 9532 \AA. We then also derive metallicities  using the \texttt{pyMCZ} code as described above, and present the results in the Appendix,  Table~\ref{tab:KECKmetallicities}. The results from the Keck spectrum are consistent with the metallicities found for the same calibrators in the integrated galaxy spectrum of the PMAS data.

\subsection{SED fit}
The host galaxy is a blue dwarf galaxy, with an observed SDSS mag of $g^\prime=18.87$ mag, and with a diameter of $\sim10\arcsec$.
We used the Le Phare code to perform SED fitting of the host galaxy of SN 2018gep using broadband data from SDSS. The physical parameters were calculated using \cite{BC2003} population synthesis models as galaxy templates. We used the photometry (corrected for the Galactic extinction of $\textnormal{A}_v=0.0286$ mag) presented in Table~\ref{tab:SEDphotometry}. 

\begin{figure}[ht!]
  % \centering 
      \includegraphics[width=\columnwidth]{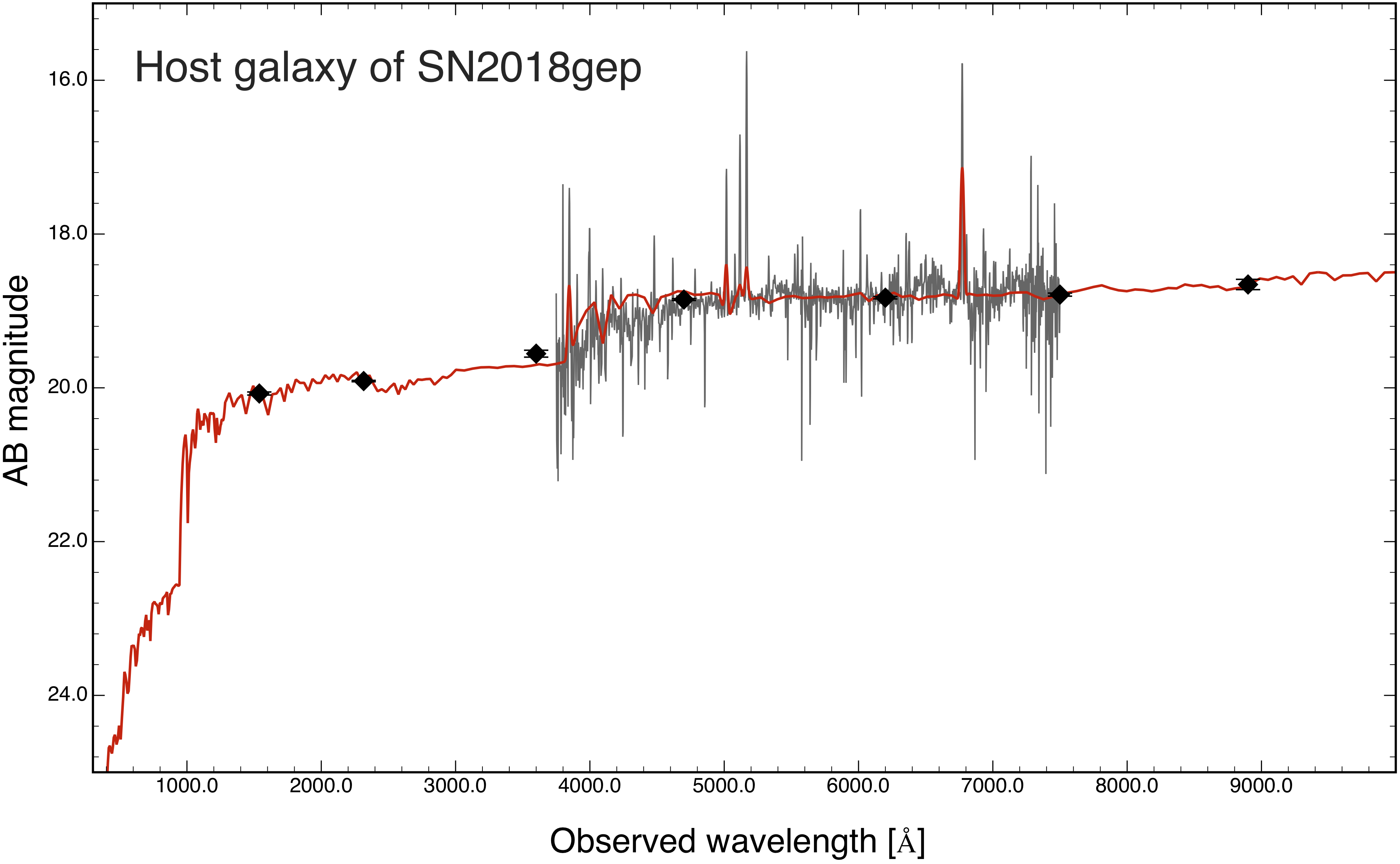}
   \caption{SED fit to the photometric data of the host galaxy of SN 2018gep (red line). We plot the spectrum of the galaxy (grey) and the photometric information for different filters (black diamonds). The plot shows the wavelength range of $300-10^4$ \AA. The SN 2018gep host-galaxy spectrum plotted in the figure was corrected for Galactic extinction and calibrated using SDSS photometry ($\textnormal{m}-\textnormal{m}_0=0.26$ mag).}
  \label{fig:SEDfit}
  \end{figure}

Our best fit has a reduced chi-square of $\sim1$ ($\chi^2=4.86$). In Figure~\ref{fig:SEDfit} we show the SED fit of the host galaxy, and the physical parameters derived are listed in Table~\ref{tab:SEDfit}. %{\bf Here goes some comment on the results}

Using this SED fitting method we infer the star-formation rate (SFR) to be SFR = $0.048^{+0.054}_{-0.010}$ [M$_\odot$\,yr$^{-1}$], while the values of the SFR based on the emission-line analysis ranges from 0.017 to 0.139 [M$_{\odot}$\ yr$^{-1}$], for the SN region ($1\arcsec\times1\arcsec$ area) and the whole galaxy, respectively. %{\color{red} MM- These values are overall consistent for the whole galaxy?  No Emission line errors}  %It is important to note that we find an intrinsic extinction of E(B-V)=0.493$\pm$0.040 for the SN region and E(B-V)=0.0 for the host galaxy in general. removed - we talk about this above
The SED reveals the total mass to be equal to M = $7.75^{+2.44}_{-1.22}$ [$10^7$\,M$_\odot$], and implies that it is a young galaxy with an age of $0.32^{+0.01}_{-0.05}$ Gyr.  

\begin{table}[ht!]
\caption{Photometry of the host of SN 2018gep used for the SED fitting.}
\label{tab:SEDphotometry} 
\centering                           
\begin{tabular}{lllllllllllllll} 
\hline\hline                
  \textbf{Filter} &\textbf{$\lambda_\mathrm{mean}$}\textbf{ [\AA]}&\textbf{mag}&\textbf{mag$_\mathrm{err}$}\\
  \hline
SDSS $u^\prime$     & 3600.0   & 19.556& 0.045 \\
SDSS $g^\prime$     & 4700.0   & 18.852& 0.012  \\
SDSS $r^\prime$     & 6200.0   & 18.828& 0.016  \\
SDSS $i^\prime$     & 7500.0   & 18.788& 0.020 \\
SDSS $z^\prime$     & 8900.0   & 18.656& 0.067  \\
GALEX NUV  & 2315.7   & 19.912& 0.009   \\
GALEX FUV  & 1538.6   & 20.074& 0.020  \\
 \hline
\end{tabular}
\end{table}

\begin{table}[ht!]
\caption{Physical parameters of the host of SN 2018gep derived using SED fitting to source photometry.}
\label{tab:SEDfit} 
\centering                           
\begin{tabular}{lllllllllllllll} 
\hline\hline                
  \textbf{Parameter [Unit]} &\textbf{Value}\\
  \hline
age [Gyr]     &     $0.32^{+0.01}_{-0.05}$     \\
M [$10^7$\,M$_\odot$]     &     $7.75^{+2.44}_{-1.22}$     \\
SFR  [M$_\odot$\,yr$^{-1}$ ]    &     $0.048^{+0.054}_{-0.010}$      \\
SSFR  [Gyr$^{-1}$]   &     $0.622^{+0.244}_{-0.043}$      \\
L$_\mathrm{NUV}$ [$10^7$\,L$_\odot$]     &     5.357   \\
L$_\mathrm{R}$ [$10^7$\,L$_\odot$]         &     5.498      \\
L$_\mathrm{K}$ [$10^7$\,L$_\odot$]         &     1.088   \\
 \hline
\end{tabular}
\end{table}

\subsection{Comparison with other SN hosts}

Most star-forming galaxies follow the fundamental mass-metallicity relationship \citep[e.g.,][]{Tremonti04} in which higher-mass galaxies also have high metallicity. Thus comparing the host galaxy of SN 2018gep to those of other transients and to the general population of star-forming galaxies as traced by the SDSS \citep{Kewley08} may give us clues about the stellar population that preferentially produces those explosions. 

\begin{figure}[ht!]
\includegraphics[width=0.49\textwidth]{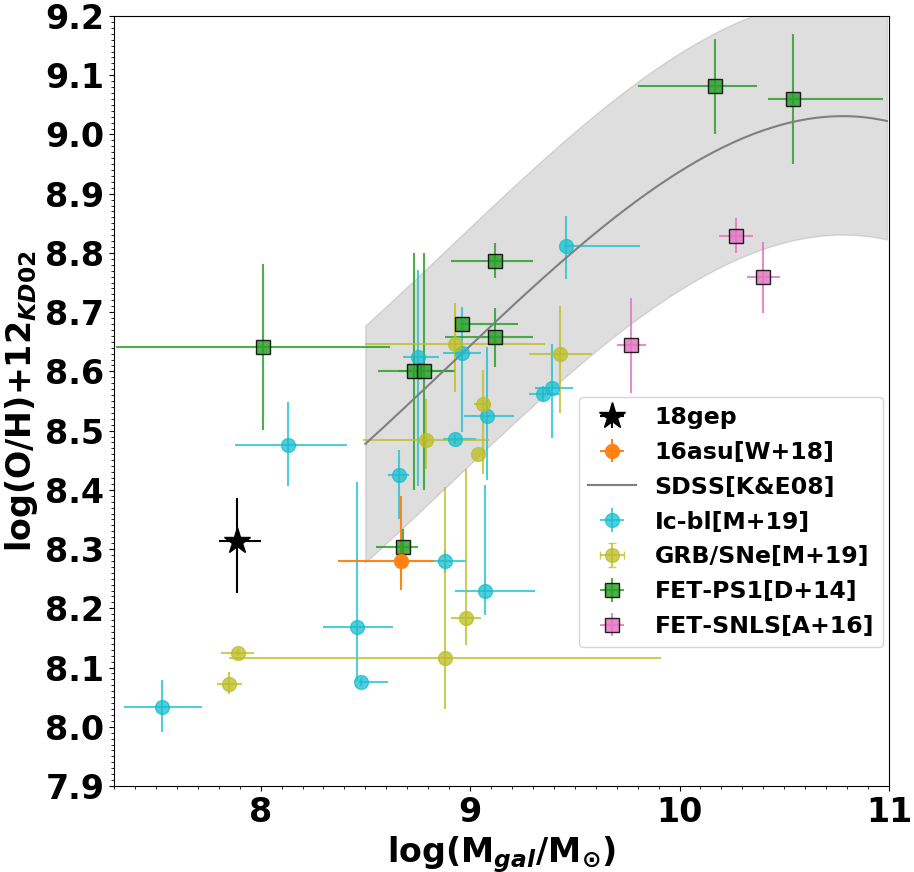}
\caption{Mass-metallicity relation of the hosts of SN 2018gep (this work) and iPTF16asu \citep{Whitesides17}, compared to the SDSS galaxy sample \citep[grey region,][]{Kewley08}, the iPTF Ic-bl SNe sample \citep{Modjaz20}, PS1 Fast Evolving Transients \citep{Drout14} and SNLS Fast Evolving Transients \citep{Arcavi16}. All values were converted to the KD02 \citep{KD02} metallicity scale using \cite{Bianco14} and published emission-line values where available, or conversion relations from \cite{Kewley08} in the remaining cases.  
\label{fig:massvmetal}}
\end{figure}

In Figure \ref{fig:massvmetal} we compare the host mass and metallicity in the KD02 \citep{KD02} scale against the values for hosts of other SNe Ic-bl, GRB-SNe and Fast Evolving Transients. The hosts of SN 2018gep and iPTF16asu are low-mass low-metallicity dwarf galaxies that lie beneath the observed SDSS population and its standard deviation \citep{Kewley08}.  The host galaxies of SN 2018gep and iPTF16asu have masses and metallicities that are broadly consistent with both the SN Ic-bl sample and the GRB-SN sample \citep[the hosts of which are also comparable to each other,][]{Modjaz20}. The host of iPTF16asu has both a mass and metallicity close to the average of these two samples while the host of SN 2018gep is on the very low mass end while having a metallicity similar to the average. Comparing the hosts of SN 2018gep and iPTF16asu with those of the fast-transient hosts, we show that their host properties are on the extreme end of the observed distribution of fast-transient hosts. The host galaxies of SN 2018gep and iPTF16asu have metallicities comparable to that of the lowest measured host from the PS1 Fast Evolving Transient sample and with the SN 2018gep host galaxy having a mass similar to the least massive and most metal-poor hosts from the PS1 sample simultaneously. In general the population of host galaxies of Fast Evolving Transients contains objects with masses and metallicities higher than those of SNe Ic-bl  or GRB-SNe.  

In Figure \ref{fig:massvmetal} the host galaxies from \citet{Pirsiainen2018} are not shown, as these galaxies had no reported metallicities. However, recent results from \cite{Wiseman20} using the host galaxies from \citet{Pirsiainen2018} have found that the host galaxy DES sample of Rapidly Evolving Transients lie in a similar space as the SNe Ic-bl \& GRB-SNe samples. The metallicity metrics used by \citet{Wiseman20} are different from those used here (PP04-O3N2 vs KD02).  Interestingly their transient sample \citep[from][]{Pirsiainen2018} does not require a strictly blue color, some of their objects are red, and for example could include objects such as PTF10vgv (See Fig. \ref{fig:lccomp}), which lacks the strong blue colors but does evolve quite rapidly. The significant, systematic offset between the host-galaxies of the PS1 sample and the DES sample likely implies either different intrinsic objects or a bias due to detection/selection method \citep{Wiseman20}; and the host of SN 2018gep is not a clear match to either of these samples.     

\section{Discussion}\label{sec:discussion}

\subsection{Comparison with Standard Models}
As we have discussed in Sections \ref{sec:lc} \& \ref{sec:spectra} SN 2018gep, while posessing the broad lines with high absorption velocities that are the defining characteristics of a SN Ic-bl, also appears  to be an outlier in the general population of  SNe Ic-bl as it exhibits an anomalous early, blue rise and is on the luminous end of the SN Ic-bl absolute magnitude distribution.  We conclude that not only is SN 2018gep different observationally than the other observed SN Ic-bl, but that it also requires a different (or at least additional) source of energy injection which is consistent with its location in Figure \ref{fig:mvr}.

\begin{figure}[]
\begin{center}
\includegraphics[width=0.4\textwidth]{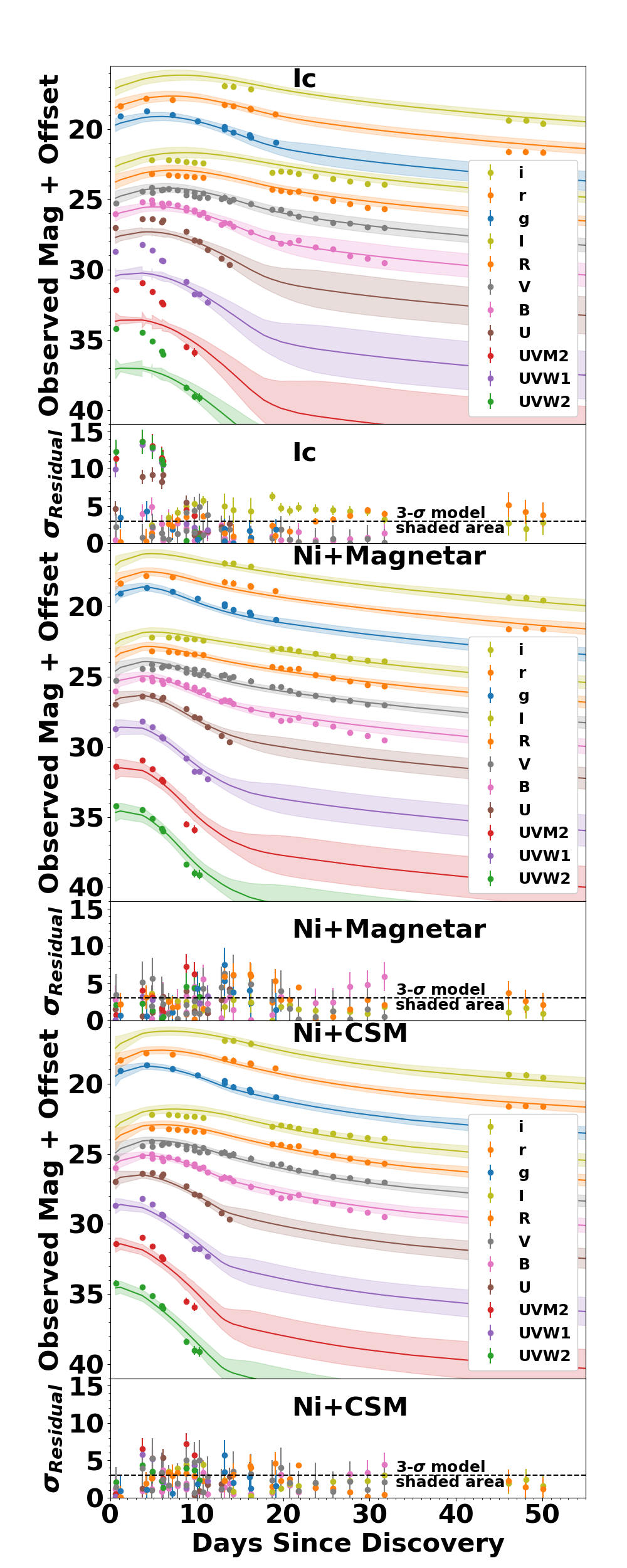}
 \caption{Simple semi-analytic model fits to the observed SN 2018gep data using the MOSFiT \citep{Guillochon18} package and NiCo decay ``Ic" \citep{Nadyozhin94}, Magnetar\citep{Nicholl17} + NiCo decay, and CSM-interraction\citep{Chatzopoulos13} + NiCo decay models.  The median (solid) and 3$\sigma$ (shaded) region of the final best-fit distribution of model data are shown with the residuals plot corresponding to the magnitude residual of the observed data scaled by the standard deviation of the models at that epoch - e.g. $(m_{obs}(t)-m_{model}(t,\theta))/\sigma_{model}(t,\theta)$ with the region below the dotted line in residuals corresponding to the shaded region of the lightcurves.  The pure Ni+Co decay model has difficulty reproducing the observed rapid, blue rise with residuals comparable to those shown in the comparison with the observed population shown in Figure \ref{fig:lccomp} (as expected of a Type Ic SNe model), while the addition of an additional power source significantly improves the fit.   \label{fig:mcomp}}

\end{center}
\end{figure}

We compare the observed SN 2018gep lightcurve with simple semi-analytic models fits using the MOSFiT package \citep{Guillochon18} in Figure \ref{fig:mcomp}. For Ic supernovae model, we see that the standard model \citep[Ni-powered explosive SNe][]{Pankey62,Arnett82,Nadyozhin94} has a difficult time reproducing the rapid, blue rise seen in the observed data.  If we add an additional source of energy injection, here Magnetar Spin-Down \citep{Kasen10,Woosley10,Nicholl17} or CSM interaction \citep{Chatzopoulos13}, we see that the early fit improves significantly. This is overall consistent with our previous conclusion that SN~2018gep is both different from a typical SN Ic-bl and most likely has an additional, or different, source of energy injection.  

The best fit model parameters for the three discussed models can be seen in Table \ref{tab:mosfit}. As these are simple semi-analytic models, the physical inference possible in such a unique case is somewhat limited.  Overall, the standard Ic model requires a significant overabundance of Ni but most closely matches the ejecta velocity and explosion date inferred from the obtained data.  The Ni + energy injection models tend to have a more realistic Ni fraction while undershooting the ejecta velocity and being on the edge of allowed explosion dates.

Of the two models with some additional non-$^{56}$Ni energy injection, the magnetar model requires a large magnetic field, B$\sim 10^{14} G$, which is comparable to that required for super-luminous SNe by similar models\citep{Nicholl17}.  While there is significant flexibility in these models, the large required value of the magnetic field most likely disfavours this energy injection method without a compelling argument for a similar compact object arising from the stellar progenitor. This would make the $^{56}$Ni+CSM interaction model the most favoured of the three models, which is consistent with the results from \citet{ho2019}. 

\begin{table}[ht!]
\caption{\label{tab:mosfit}Best-fit model parameters in the MOSFIT package for powering the UV-optical light curves of SN~2018gep (Fig. \ref{fig:mcomp})}
\centering                           
\begin{tabular}{llll} 
\hline\hline                
  \textbf{Parameter[Unit]}&\textbf{Ic}&\textbf{Ni+Mag}&\textbf{Ni+CSM}\\
  \hline

log M$_{ej}$[M$_\odot$] & $-0.12_{-0.04}^{+0.03}$ & $-0.58_{-0.06}^{+0.08}$ & $-0.31_{-0.30}^{+0.19}$ \\
log f$_{Ni}$ & $-0.01_{-0.01}^{+0.002}$ & $-0.30_{-0.17}^{+0.17}$ & $-0.56_{-0.25}^{+0.32}$ \\
t$_{exp}$[days] & $-5.42_{-0.74}^{+0.64}$ & $-2.4_{-0.75}^{+0.53}$ & $-1.9_{-0.67}^{+0.50}$ \\
log v$_{ej}$[km s$^{-1}$] & $4.50_{-0.06}^{+0.04}$ & $4.45_{-0.70}^{+0.4}$ & $4.7_{-0.27}^{+0.21}$ \\
log $\kappa$[cm$^2$ g$^{-1}$] & $-0.98_{-0.02}^{+0.03}$ &  $-0.15_{-0.18}^{+0.23}$ & $ -0.24_{-0.33}^{+0.33}$\\
log n$_{\rm H,host}$    & $17.66_{-1.13}^{+0.93}$   & $20.49_{-1.92}^{+0.53}$ & $17.94_{-1.40}^{+1.52}$\\
log $\sigma$            & $-0.08_{-0.02}^{+0.03}$   & $-0.26_{-0.03}^{+0.03}$ & $-0.27_{-0.02}^{+0.03}$\\
log T$_{\rm min}$ (K)   & $3.63_{-0.03}^{+0.03}$    &  & $3.75_{-0.02}^{+0.02}$ \\
log B                   &   & $0.97_{0.04}^{+0.02}$                          & \\
M$_{\rm NS}$(M$_\odot$) &   & $1.04_{-0.03}^{+0.07}$                         & \\
P$_{\rm spin}$ (ms)     &   & $8.16_{-3.02}^{+1.43}$                         & \\
$\theta_{\rm PB}$ (rad) &   & $1.34_{-0.17}^{+0.15}$                         & \\
log M$_{\rm CSM}$       &   &   & $-0.97_{-0.02}^{+0.04}$ \\
log $\rho$              &   &   & $-11.27_{-0.06}^{+0.06}$ \\
\hline
\end{tabular}
\tablecomments{Best fit values and 2-$\sigma$ errors for model parameters.  See \citet{Guillochon18} \& \citet{Chatzopoulos13, Nicholl17, Nadyozhin94} for parameter details.}
\end{table}

\subsection{Pre-Explosion Variability, CSM Interaction, and Comparison with Other Work}
The work done by \citet{ho2019} on SN~2018gep shows the detection of pre-explosion variability and inferred mass-loss by the progenitor star and the subsequent interaction between the pre-explosion ejected mass and the supernovae shock.  This is a well-substantiated \& physically motivated model for SN~2018gep that is overall consistent with our more general \& data-driven finding of some additional source of energy injection to be present early on in the light curve.

The similarity between SN~2018gep and the PS-1 and DES `Fast Evolving Transients' while also noted by \citet{ho2019}, is not studied in significant detail by them as we do here including our light-curve and environment studies and folding iPTF16asu into this as well.  We find some similarity to both SLSNe (though SN2018gep has an even bluer spectrum pre-max than SLSNe) and GRB-SNe (in the light curve and spectra) which is consistent with the \citet{ho2019} findings of potential SLSNe spectral features and the high velocities only seen otherwise in GRB-SNe.  %\citet{Arcavi16} also find that one of the `Rapdily Evolving Luminous Transients' in the supernovae legacy sample has some similarity to the  ultra-long GRB-SN 2011kl, which 

\section{Conclusion - Fast Blue Optical Transients, SN Ic-bl, or Both?}
SN 2018gep is a SN Ic-bl with anomalously blue colors ($\gtrsim 4$ mag in UVW2$-$v or $\sim 2$ mag in g$-$r) at early epochs and a rapid rise time (t$_{rise} =6.2 \pm 0.8$ days). This anomalous behavior is also seen in its early, blue, nearly featureless spectrum, which at later times (after maximum light) shows more significant absorption lines while maintaining its atypical blue continuum.  With a host metallicity of log(O/H)$+12 = 8.31^{+0.07}_{-0.09}$ (from the SN region)  and host galaxy mass of M$_{host}= 7.8^{+2.4}_{-1.2} \times 10^7$\,M$_\odot$, it is within the typically observed range of SN Ic-bl host parameters and on the edge of the FBOT host property distribution. All these properties place SN 2018gep as a significant outlier when compared with other SNe Ic-bl except for iPTF16asu, while at the same time it is on the edge of the observed parameter space for FBOTs.  In addition to these derived properties, its general photometric evolution occurs in a highly similar manner to the observed PS1 FBOTs \citep[PanSTARRS,][]{Drout14}, which is the only FBOT sample with well observed rise times.  When compared with  simple analytical SN Ic models, we see that the standard SN Ic model has difficulty reproducing the rapid blue rise while the post-peak data is more well matched by the models. We find that an additional energy-injection mechanism (here, CSM interaction or magnetar coupling) improves the early time fit significantly.  

The observations of SN 2018gep highlight the time (and to a lesser extent sensitivity) dependant nature of our classification schemes for these mysterious transients.  If we had poorer quality observations of iPTF16asu \& SN 2018gep we would have likely called  these events just FBOTs given their blue colors, rapid rises and nearly featureless blue spectra before and around maximum light.  However, if we had only obtained late observations (or had fewer colors) we would have likely classified SN 2018gep as a more standard SN Ic-bl given that its later spectra and colors are more closely matched to the broader SN Ic-bl sample and that the later light curve is well fit by the typical models.  In fact, if only red data (i.e., rest-frame $g^\prime$-band filter and red-wards) had been obtained, as is common in many transient surveys, this SN would have looked much more similar to the SN Ic-bl sample as a whole and the generic analytical $^{56}{\rm Ni}$ driven model would have produced a reasonable fit to the data. Similarly, if the early emission had been missed (e.g. t$<$10 days after discovery), this object would have appeared more like a typical SN Ic-bl.  This object highlights the need for missions such as {\em Swift}\citep{Gehrels04} and the proposed  Gravitational-wave Ultraviolet Counterpart Imager (GUCI) Network\citep{Cenko19}, which enable the prompt UV observations crucial for classification as well as our understanding of the atypical explosion and energy injection mechanisms of transient events like this.

However, with a fortuitous object that is bright, nearby, and discovered promptly - such as SN 2018gep - it is possible to acquire a detailed data set including: early time data with high cadence and colors, multi-wavelength information, a spectral time series and host galaxy observations - all of which we present here. It is only this more complete data set that illustrates the SN transitioning from a rapidly rising blue transient to a SN Ic-bl, and this photometric and spectroscopic evolution may provide some insight into other observed FBOTs and extreme SNe Ic-bl.  

When compared against the PS1-FBOT sample (the only such with host information and measured rise times), both SN 2018gep and iPTF16sau show a similar photometric rise and decline time.  While the color data are noisy, due to the simplistic comparison across red-shifts performed with minimum assumptions in addition to the intrinsic variability of the observed FBOT sample, the observed results for both of these objects lie well within the observed PS1-FBOT distribution. The host environments of SN 2018gep and iPTF16asu occupy a similar region of the host galaxy mass vs. metallicity distribution as the other SN Ic-bl \& GRB/SNe from \citet{Modjaz20} and are on the edge of the observed Fast Evolving Transient hosts phase space.  %When compared with simple analytic models, the data suggests that the Ic model is a reasonable fit at late times and in the redder bands, but requires an additional source of energy injection to fit the rapid rise and anamolous blue color.  

Not all of the observed FBOTs (or even all FBOTs in only the PS1 sample) can be like SN 2018gep or iPTF16asu.  The observed FBOTs span too broad a range of host environments and intrinsic magnitudes to be consistent with the general SN Ic-bl and SNe-GRB sample.  Furthermore, while many FBOTS have similar photometric evolution, there are notable exceptions, such as AT2018cow with its rapid evolution but minimal color evolution and one object in the PS1 sample with emission on longer timescales. Furthermore, \citet{Arcavi16} compare a number of power sources and conclude that from their samples not all similar events can be powered by the same source. There is a need for significantly more multi-epoch spectra across FBOTS as a whole, as we cannot make strong conclusions without a greater sample of significantly pre- and post-peak spectra.  

However, we speculate that if the physical explosion of SN~2018gep and iPTF16asu is that of a SN Ic-bl with a rapid, blue rise driven by an additional source of energy injection, then perhaps the FBOTs with similar photometric evolution (e.g. most of the PS1 sample and many others) could share a similar explosion or energy injection mechanism.  It could be that this energy-injection mechanism drives the observed early, blue rise common to the sample, but with differing progenitor stars (and underlying supernovae) that may lead to much of the observed variance in the sample.  

This model - a variety of underlying explosions with an additional source of early, blue emission -  would be consistent with the reports of pre-explosion variability and a CSM interaction driven model by \cite{ho2019}, and perhaps one diagnostic of this common FBOT energy injection mechanism might be a systematic search for pre-explosion variability across a larger sample of well studied FBOT SNe. While historically difficult to do, the increasing cadence and depth of large area synoptic surveys is making this increasingly feasible.  
In the future, the Vera Rubin Observatory LSST will be able to fortuitously provide pre-explosion images throughout the survey's 10 year duration, enabling the search for signatures of a common energy injection mechanism.  Another key to further understand the nature of these events will be the acquisition of multi-epoch spectroscopy for a significant sample size of fast evolving transients. Time series spectra allow us to test our hypothesis whether, as a sample, these objects develop significant variations at later times from their featureless blue continuum around maximum light, and if they evolve similarly or with significant diversity.  Additional UV observations (whether from {\em Swift}, GUCI, or another mission) will similarly be key as the modestly blue optical colors as seen in SN 2018gep  belied a significantly greater UV flux; and understanding how common and energetic this blue emission is will allow us to constrain the explosion mechanism and progenitor further.  

\acknowledgments

This research made use of Astropy,\footnote{http://www.astropy.org} a community-developed core Python package for Astronomy \citep{astropy1, astropy2}.

Some of the data presented herein were obtained at the W. M. Keck Observatory, which is operated as a scientific partnership among the California Institute of Technology, the University of California and the National Aeronautics and Space Administration. The Observatory was made possible by the generous financial support of the W. M. Keck Foundation.
The authors wish to recognize and acknowledge the very significant cultural role and reverence that the summit of Maunakea has always had within the indigenous Hawaiian community.  We are most fortunate to have the opportunity to conduct observations from this mountain.

M.M. and the SNYU group are supported by the NSF CAREER award AST-1352405, by the NSF award AST-1413260 and by a Humboldt Faculty Fellowship. MM is grateful for her sabbatical stay supported by the Center for Computational Astrophysics at the Flatiron institute where some of this work was accomplished.

KB acknowledges financial support from the Ministerio de Economía y Competitividad through the Spanish grant BES-2014-069767. KB, CT and AdUP acknowledge support from the Spanish research project AYA2017-89384-P. CT acknowledges support from funding associated to a Ram\'on y Cajal fellowship RyC-2012-09984. AdUP acknowledges support from funding associated to a Ram\'on y Cajal fellowship RyC-2012-09975. LI acknowledges support from funding associated to a Juan de la Cierva Incorporaci\'on fellowship IJCI-2016-30940. DAK acknowledges support from the Spanish research projects AYA 2014-58381-P, AYA2017-89384-P, from Juan de la Cierva Incorporaci\'on fellowship IJCI-2015-26153, and from Spanish National Project research project RTI2018-098104-J-I00 (GRBPhot).

J.V. and his research group at Konkoly Observatory is supported by the ``Transient Astrophysical Objects'' GINOP 2.3.2-15-2016-00033 project of the National Research, Development and Innovation Office (NKFIH), Hungary, funded by the European Union.   KV and LK thank the financial support from the National Research, Development and Innovation Office (NKFIH), Hungary, under grants NKFI-K-131508 and NKFI-KH-130526. AB and KV are supported by the Lend\"ulet program grant LP2018-7/2019 of the Hungarian Academy of Sciences. 
TND also acknowledges the support of the Hungarian OTKA Grant No. 119993.
%DAK acknowledges support from the Spanish research
%projects AYA 2014-58381-P, AYA2017-89384-P, from Juan de la Cierva
%Incorporaci\'on fellowship IJCI-2015-26153, and from Spanish research
%project RTI2018-098104-J-I00 (GRBPhot)

L.G. was funded by the European Union's Horizon 2020 research and innovation programme under the Marie Sk\l{}odowska-Curie grant agreement No. 839090. This work has been partially supported by the Spanish grant PGC2018-095317-B-C21 within the European Funds for Regional Development (FEDER).

R.G.B. acknowledges financial support from the Spanish Ministry of
Economy and Competitiveness through grant AYA2016-77846-P and from the
State Agency for Research of the Spanish MCIU through the ``Center of
Excellence Severo Ochoa" award to the Instituto de Astrof\'isica de Andaluc\'ia (SEV-2017-0709).

These observations made use of the LCO network.  DAH, CP, DH, and JB are supported by NSF Grant AST-1911225 and NASA Grant 80NSSC19k1639

\facility{Keck:I (LRIS), Sloan, Konkoly:Schmidt}

\bibliographystyle{aasjournal}
\bibliography{main}


%%\vspace{5mm}
%%\facilities{HST(STIS), Swift(XRT and UVOT), AAVSO, CTIO:1.3m,


%%\software{astropy \citep{2013A&A...558A..33A},  
%%          Cloudy \citep{2013RMxAA..49..137F}, 
%%          SExtractor \citep{1996A&AS..117..393B}
%%          }



\appendix

\section{Host Galaxy Analysis Details: IFU and long-slit spectra}\label{IFU}
\begin{figure*}[ht!]
\centering  
\includegraphics[width=6.5cm]{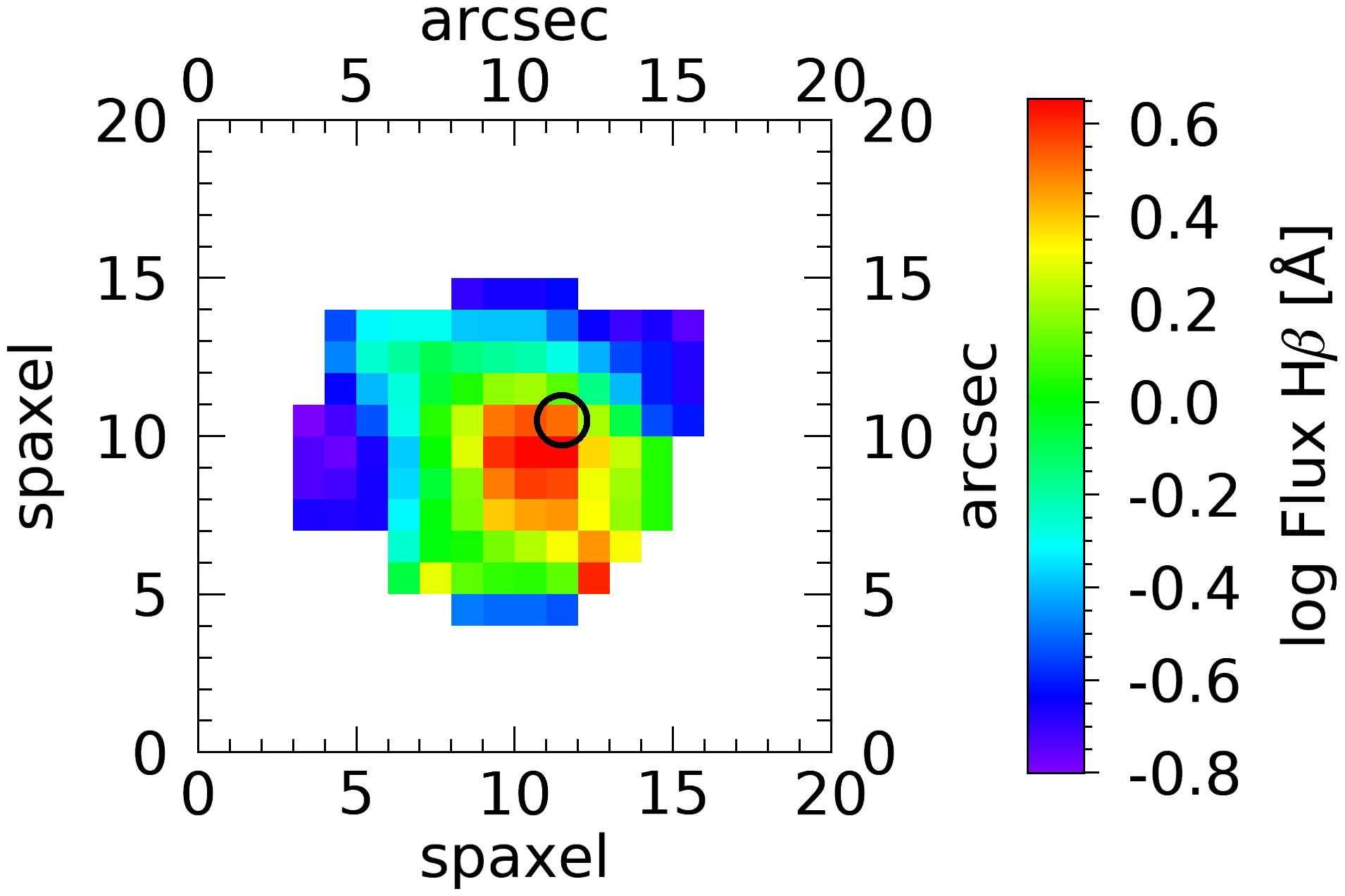}
\includegraphics[width=6.5cm]{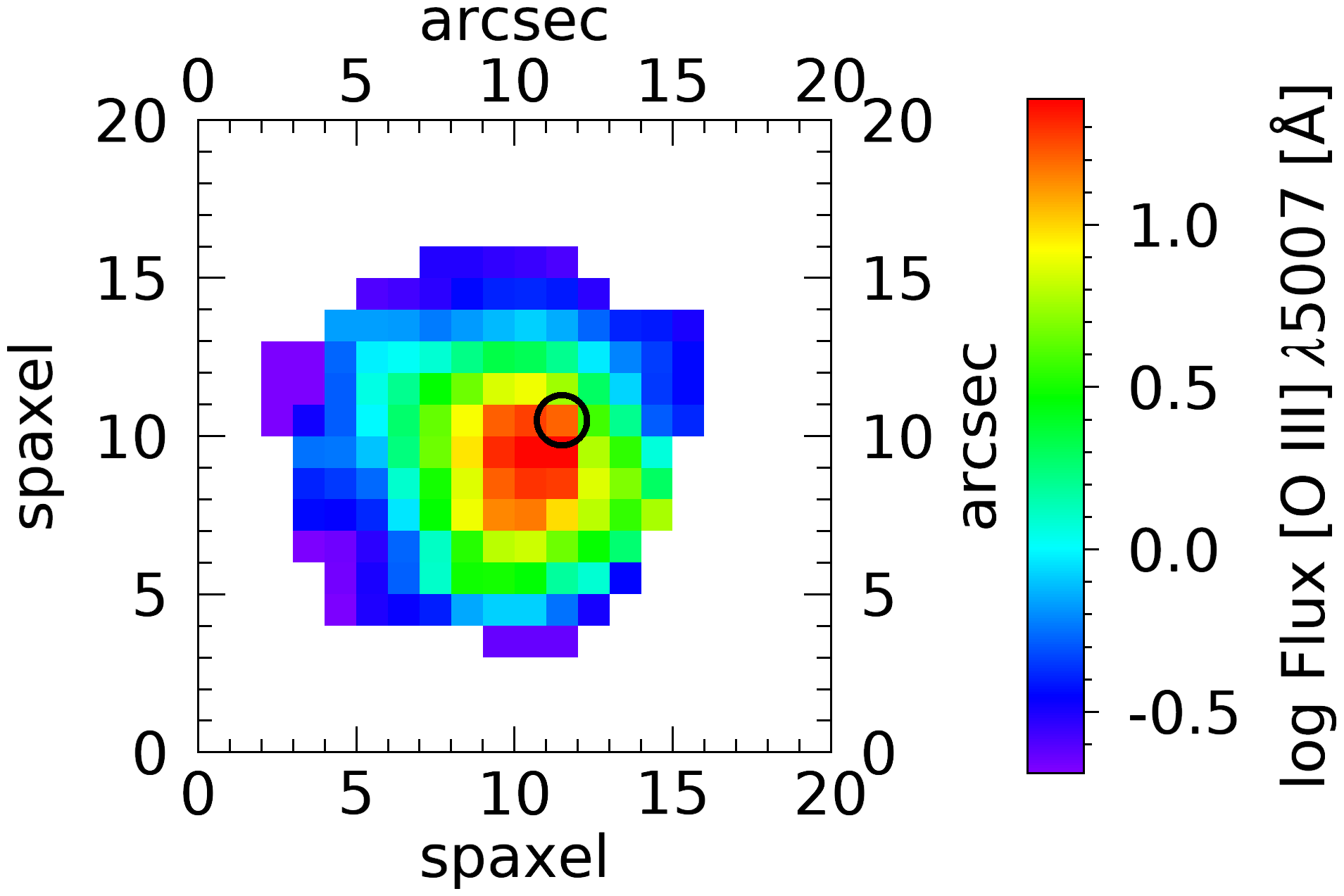}\\
\includegraphics[width=6.5cm]{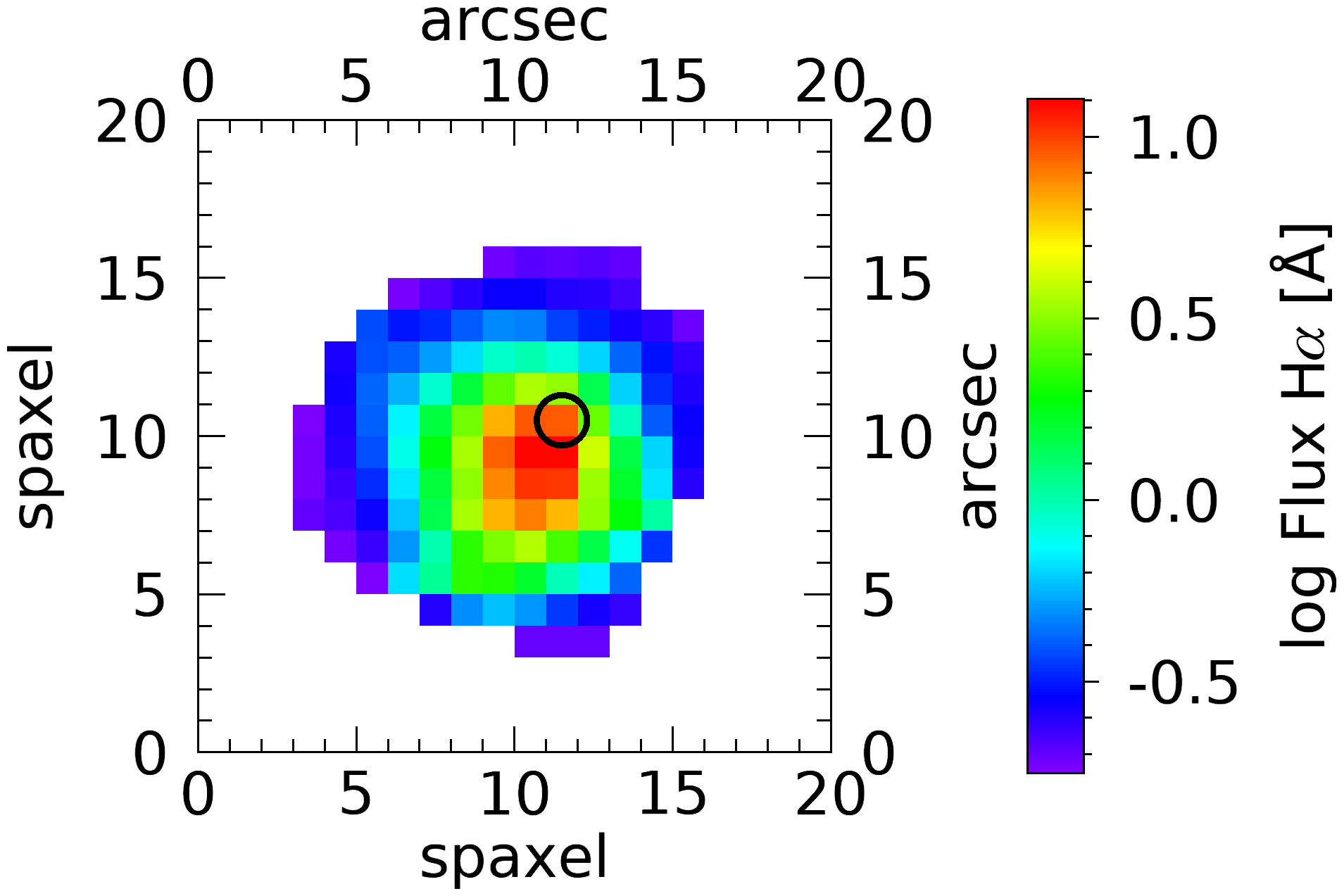}
\includegraphics[width=6.5cm]{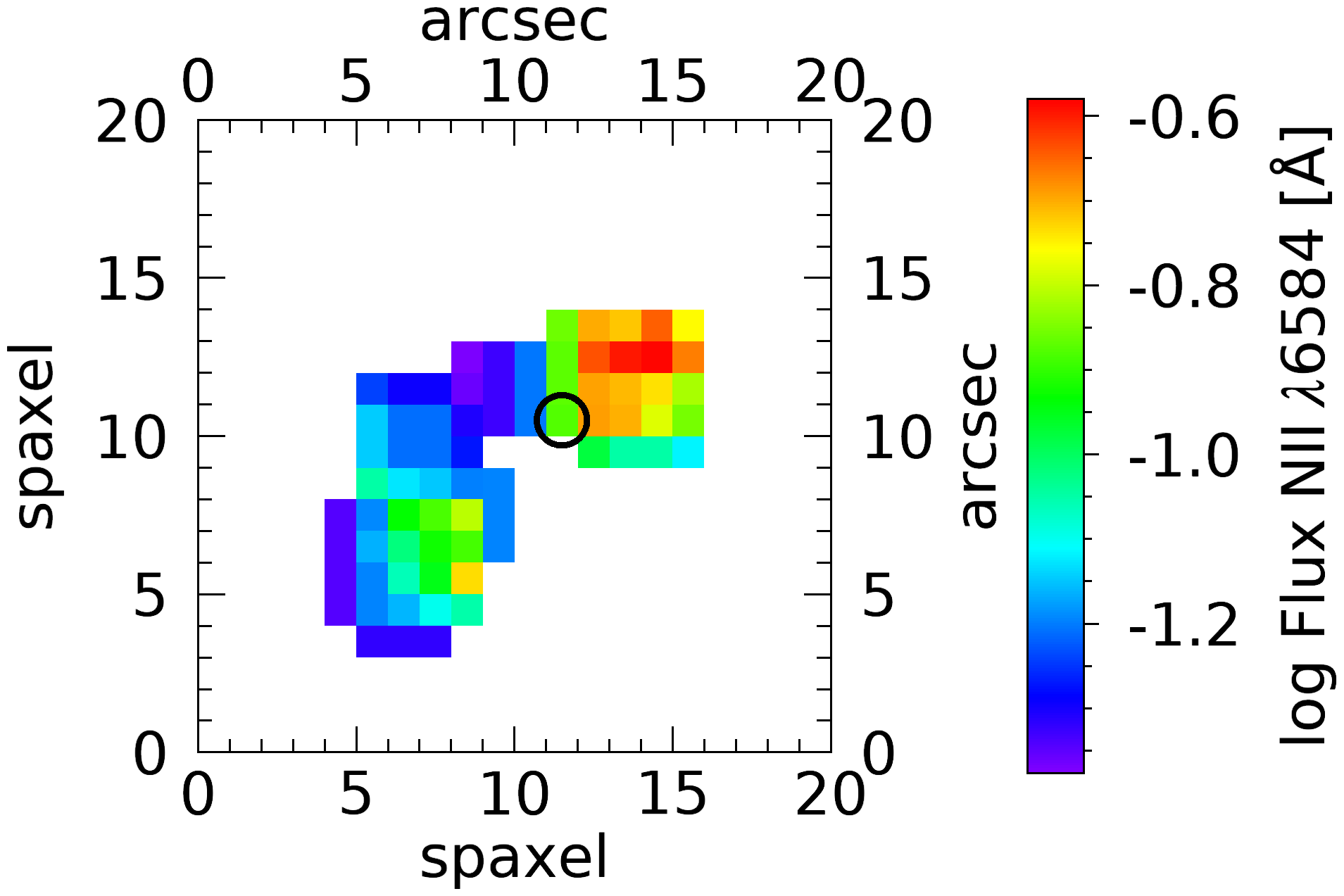}
\caption{Distribution map of the emission-line fluxes in the SN 2018gep host galaxy: H$\beta$%, [O III] $\lambda$ 4959\AA%
, [O III] $\lambda$ 5007\AA, H$\alpha$, [N II] $\lambda$ 6584\AA. The black circle indicates the position of SN 2018gep.}
\label{fig:linemaps}
\end{figure*}

\begin{figure*}[ht!]
\centering  
\includegraphics[width=5.6cm]{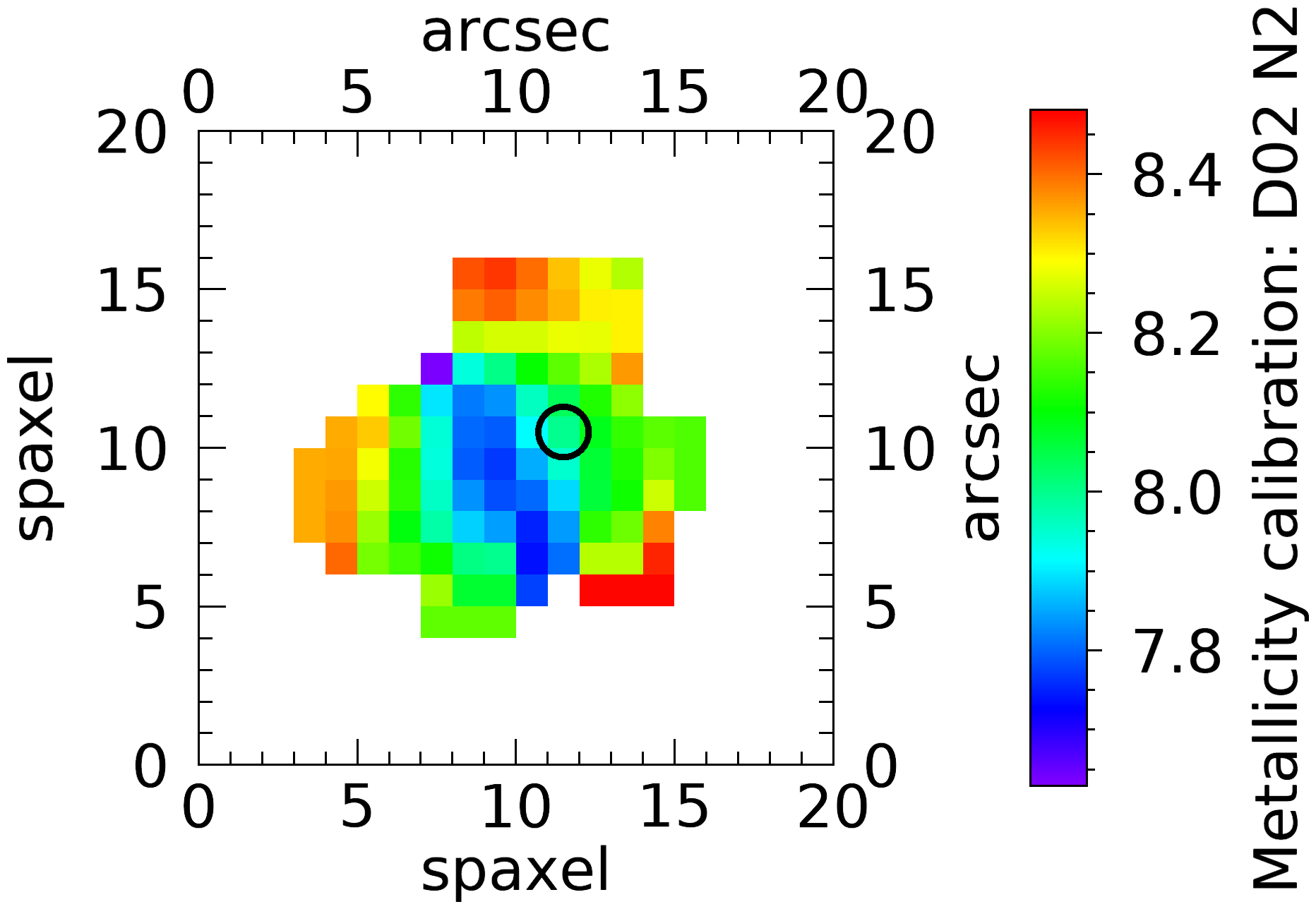}
\includegraphics[width=5.6cm]{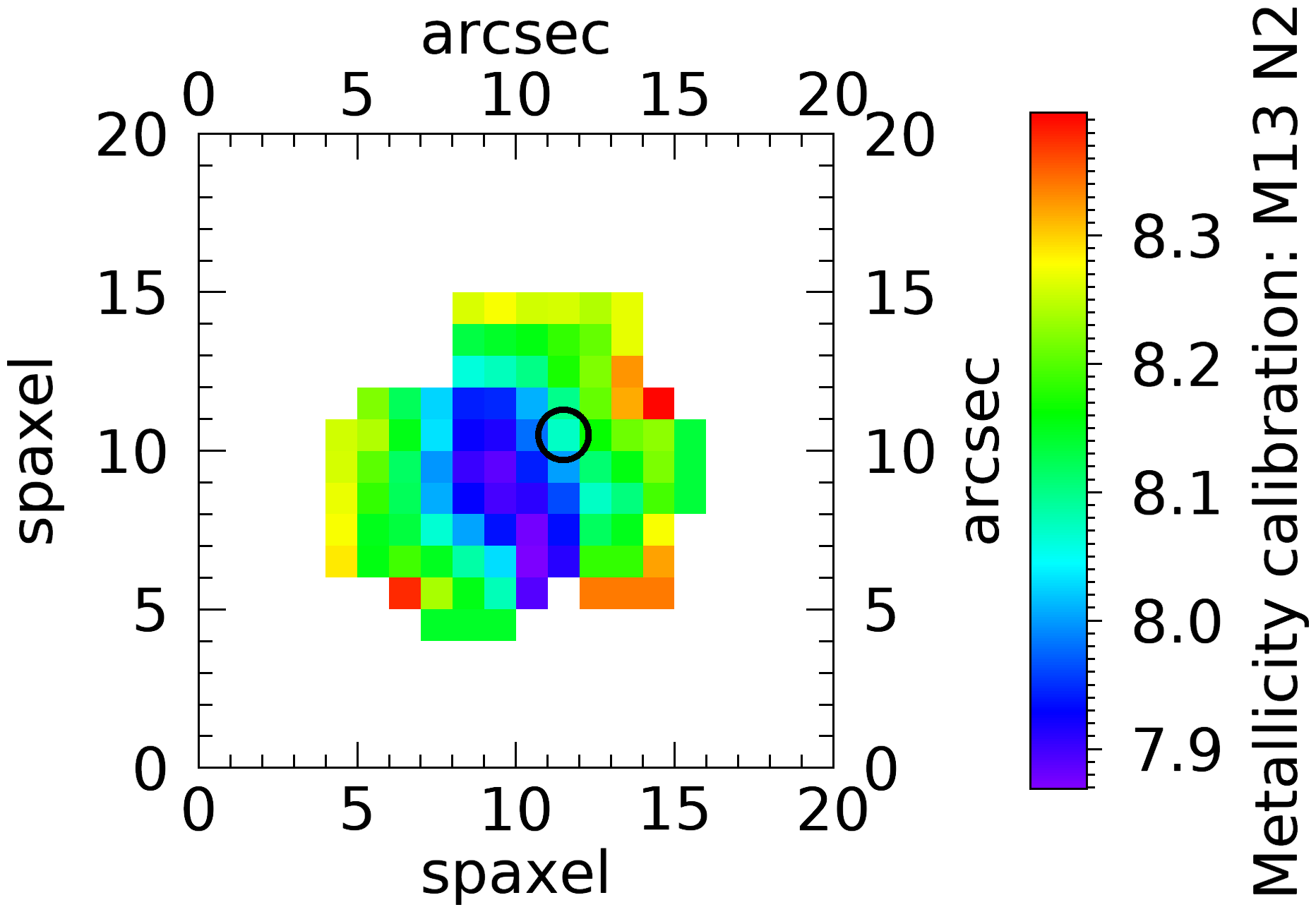}
\includegraphics[width=5.6cm]{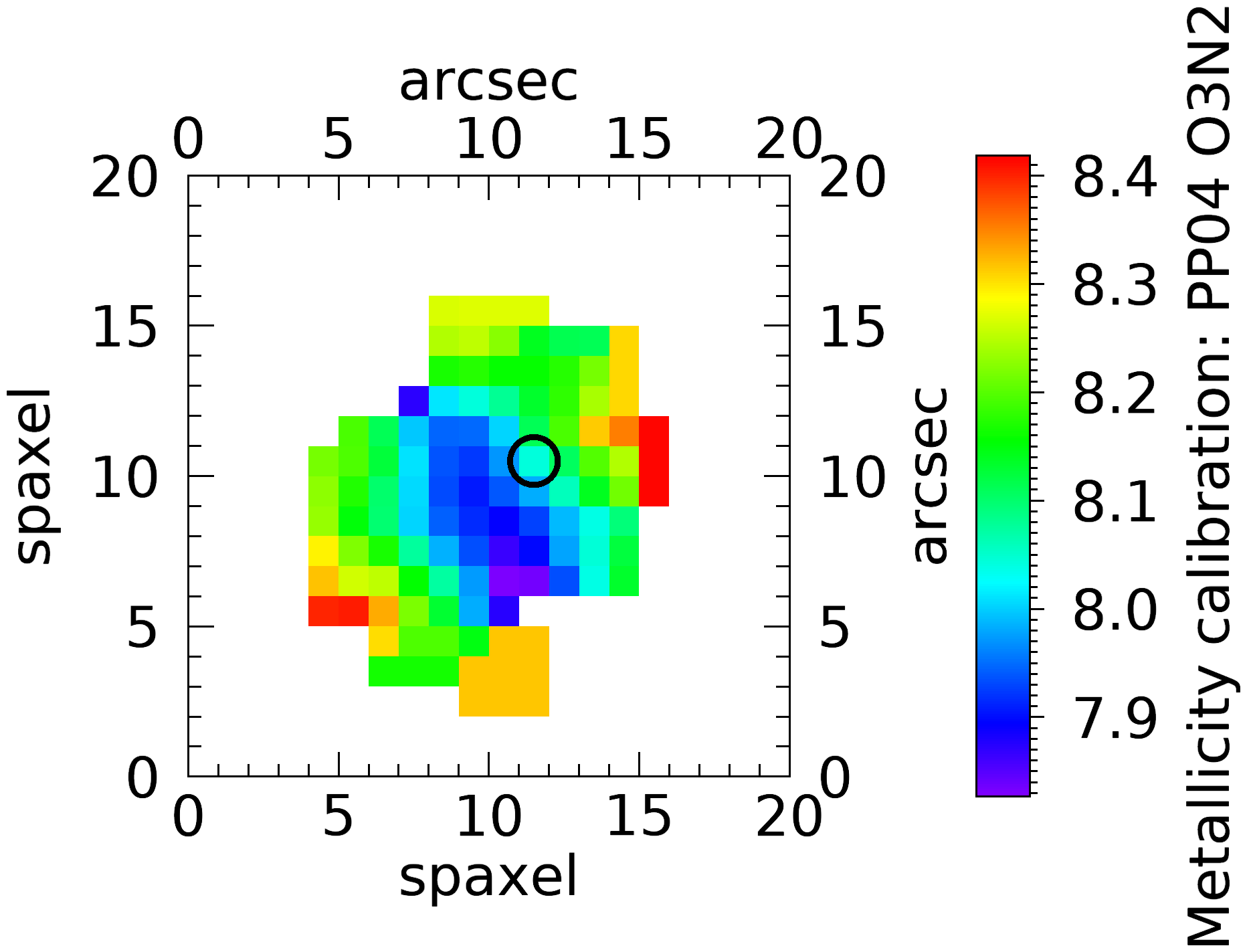}
\caption{The maps of the SN 2018gep host-galaxy metallicities derived with different metallicity calibrations: using the N2 parameter from the \cite{KD02} calibration (left), N2 parameter in the  \cite{M13O3N2} calibration (middle) and using the O3N2 parameter in the calibration of \cite{PP04} (right). The black circle indicates the position of SN 2018gep.}
\label{fig:metallicitymaps}
\end{figure*}

Supplementary information to the data reduction and analysis described in Section \ref{sec:spectra}.  In Figure \ref{fig:linemaps} we present host galaxy maps of commonly used emission lines.  In Figure \ref{fig:metallicitymaps} we present maps of the derived host galaxy metallicities for a selection of calibrators.  In Table \ref{tab:KECKfluxes} we present emission line fluxes from the longslit LRIS spectrum obtained from the W.M. Keck Telescope.  In Table \ref{tab:KECKmetallicities} we present the derived metallicities using the emission lines from the Keck spectrum for a variety of calibrators.  In general we find reasonable agreement between the properties derived from the Keck Spectrum and spatially average properties of the IFU data.  

%{\color{magenta} need some text here} {CT: not sure we need much text here... we can just reference the figures and emission line flux table in the main text refering to appendix plots/tables since all the necessary explanation on the data, reduction, analysis and results is now in the main text.}

\vspace{5cm}

\begin{table}[ht!] 
\caption{The emission-line fluxes from the LRIS long-slit spectrum with galactic extinction correction applied. All fluxes are in 10$^{-16}$  erg s$^{-1}$ cm$^{-2}$.}
\label{tab:KECKfluxes} 
\centering                           
\begin{scriptsize}
\begin{tabular}{lllllllllllllll} 
\hline\hline                
  \textbf{Emission line} &\textbf{$\lambda$ [\AA]}  & \textbf{Host galaxy}  \\
 \hline
$[$OII$]$ $\lambda$3727\AA     &  3726.5000 &    35.57$\pm$ 0.7475 \\
H$\beta$ $\lambda$4861\AA      &  4859.3460 &    21.13$\pm$ 0.3292 \\
 $[$OIII$]$ $\lambda$4959\AA    &  4956.8080 &    35.10$\pm$ 0.3747 \\
$[$OIII$]$ $\lambda$5007\AA    &  5004.7220 &    104.0$\pm$ 0.6245 \\
$[$OI$]$ $\lambda$6300\AA      &  6298.3150 &   0.668$\pm$0.0603 \\
H$\alpha$ $\lambda$6563\AA     &  6561.6660 &    56.62$\pm$ 0.3015 \\
$[$NII$]$ $\lambda$6584\AA     &  6582.9170 &    1.547$\pm$0.05567 \\
$[$SII$]$ $\lambda$6717\AA     &  6715.4190 &    3.384$\pm$0.08153 \\
$[$SII$]$ $\lambda$6731\AA     &  6729.7800 &    2.506$\pm$0.07482 \\  
$[$SIII$]$ $\lambda$9069\AA    &  9068.5210 &    2.503$\pm$0.04833 \\
$[$SIII$]$ $\lambda$9532\AA    &  9530.5660 &    7.763$\pm$ 0.1247 \\ 
 \hline
\end{tabular}
\end{scriptsize}
\end{table}

\begin{table}[ht!]
\caption{Derived Oxygen Abundance based on the LRIS long-slit spectrum in different scales using the code from \citet{Bianco2016}}
\label{tab:KECKmetallicities} 
\centering                           
\begin{scriptsize}
\begin{tabular}{c c c c c c c c c} 
\hline\hline                
\textbf{Calibrator}	& \textbf{Host galaxy}   \\
\hline%newresults
D02                 &         7.979   +  0.157    -  0.166\\
Z94                 &         8.440   +  0.004    -  0.003\\
M91                 &         8.077   +  0.015    -  0.015\\
PP04 N2Ha           &         8.053   +  0.007    -  0.007\\
PP04 O3N2           &         8.008   +  0.006    -  0.005\\
P10 ONS             &         8.933   +  0.025    -  0.025\\
P10 ON              &         7.888   +  0.035    -  0.035\\
M08 N2Ha            &         8.033   +  0.015    -  0.015\\
M08 O3O2            &         8.059   +  0.009    -  0.010\\
M13 N2              &         8.018   +  0.046    -  0.045\\
KD02 N2O2           &         7.601   +  0.035    -  0.031\\
KK04 N2Ha           &         8.250   +  0.015    -  0.016\\
KK04 R23            &         8.286   +  0.012    -  0.013\\
KD02comb            &         8.182   +  0.014    -  0.014\\
\hline
\end{tabular}      
\end{scriptsize}
\end{table}

\end{document}